\documentclass{aa}

\usepackage{graphicx}
\usepackage{txfonts}

\newcommand{\Mpc}{$h^{-1}$\thinspace Mpc}

\newcommand{\etal}{{\rm et al.~}}

\newcommand{\be}{\begin{equation}}
\newcommand{\ee}{\end{equation}}

\def\apjl{ApJL}
\def\apjs{ApJS}

\edef\aap{A\&A}

\begin{document}   

\title{ The richest superclusters} 
\subtitle{II. Galaxy populations} 

\author{ M. Einasto\inst{1} \and E. Saar\inst{1} 
\and J. Einasto\inst{1} 
\and E. Tago\inst{1} \and L. J. Liivam\"agi\inst{1} \and
V.J. Mart\'{\i}nez$^{2}$ \and J.-L. Starck$^{3}$ 
\and V. M\"uller\inst{4}  \and P. Hein\"am\"aki\inst{5}  
\and P. Nurmi\inst{5} \and M. Gramann\inst{1}  
\and  G. H\"utsi\inst{1}
}

\institute{Tartu Observatory, EE-61602 T\~oravere, Estonia
\and
Observatori Astron\`omic, Universitat de Val\`encia, Apartat
de Correus 22085, E-46071 Val\`encia, Spain 
\and 
CEA-Saclay, DAPNIA/SEDI-SAP, Service d'Astrophysique, F-91191 Gif
sur Yvette, France 
\and
Astrophysical Institute Potsdam, An der Sternwarte 16,
D-14482 Potsdam, Germany
\and
Turku University,
Tuorla Observatory, V\"ais\"al\"antie 20, Piikki\"o, Finland 
}

\date{ Received 2007; accepted} 

\authorrunning{M. Einasto et al.}

\titlerunning{2dF rich superclusters}

\offprints{M. Einasto }

\abstract 
{}
{ We study the morphology of galaxy populations of the richest superclusters
  from the catalogue of superclusters of galaxies in the 2dF Galaxy Redshift
  Survey.  }
{ We use the luminosity density distribution and Minkowski functional $V_3$ 
  to study substructures
  in superclusters as formed by different populations of galaxies.  We
  compare the properties of grouped and isolated
  galaxies in regions of different density in superclusters.  }
{ In high-density cores of rich
  superclusters there is an excess of early type, passive galaxies, among
  galaxies in groups and clusters, as well as among those which do not belong
  to groups, while in lower density outer regions there are more blue,
  star-forming galaxies both in groups and among those galaxies which do not
  belong to groups. This also shows that the galaxy content of groups depends
  on the environment where the groups reside in.
  The density distributions and the behaviour of the Minkowski functional 
  $V_3$ for different superclusters show that substructures
  in superclusters as traced by different populations of galaxies are very
  different. 
   }
{ We show how the Minkowski functional $V_3$ can be used to study substructures
  in superclusters. Both local (group/cluster) and global (supercluster)
  environments are important in forming galaxy morphologies and star formation
  activity.  Our study shows large differences between individual very rich
  superclusters, which cannot be solely due to selection effects. They indicate
  that there are differences in their 
  formation times and evolution stages.  The presence
  of a high density core with X-ray clusters and a relatively small fraction
  of star-forming galaxies in the supercluster SCL126 may be an indication
  that this supercluster has started its evolution earlier than 
  the supercluster SCL9.  }


\maketitle

\section{Introduction}

{\scriptsize
\begin{table*}[ht]
\caption{Data on rich superclusters }
%\tiny
\begin{tabular}{lrrrrrrrrrrcc} 
\hline 
 ID & R.A. & Dec & Dist & $N_{gal}$ &$M_{lim}$ & $N_{vol}$ & $N_{cl}$ &$N_{gr}$ &$N_{ACO}$&$N_X$ & $\delta_m$ &   $L_{tot}$ \\
     & deg & deg & \Mpc  &       &   &&  &  &     &   &              \\
\hline 
SCL126 (152)& 194.71 & -1.74& 251.2  &  3591  & -19.25&1308 &18 &  40,2 &  9     &  4       & 7.7 &  0.378E+14 \\
SCL10 (5)   &   1.85& -28.06& 177.4  &   952  & -17.50& 757 & 5 &   5   & 1 (19) &  (5)   & 6.2 &   0.482E+13 \\
SCL9 (34)   &   9.85& -28.94& 326.3  &  3175  & -19.50&1176 &24 &  26,9 &  12 (25) &  2(6) & 8.1 &   0.497E+14\\
\label{tab:1}     
\end{tabular} 

Note: 
Identity ID after Einasto et al. (2001) with the name of Paper I in
parenthesis; with sky coordinates and distance $D$ for our cosmology; 
the galaxy number $N_{gal}$ for the whole superclusters,
and magnitude limit $M_{lim}$ and the galaxy number $N_{vol}$ for 
volume limited superclusters; 
$N_{cl}$ and $N_{gr}$ are density field 
cluster and group numbers according to paper I;  
$N_{ACO}$ gives the number of Abell clusters in this part of
the supercluster that is covered by 2dF survey; the number inside  
parenthesis is the total number of Abell clusters in this
supercluster, by Einasto et al. (2001) list;
$N_X$ -- number of X-ray clusters;
$\delta_m$ -- the mean values of the luminosity density field 
in superclusters, in units of mean density;
$L_{tot}$ -- supercluster total luminosity in Solar units.
\end{table*}            
}

The present series of papers is devoted to the detailed study of rich 
superclusters. 
In the first paper (Einasto et al., 
\cite{e07d}, hereafter RI) we studied the morphology of rich 
superclusters using Minkowski functionals, and showed that typically, 
the shapes of rich superclusters can be characterized by a multi-branching 
filament.  In the present paper we shall study galaxy populations in 
individual rich superclusters.  This study continues  our studies of the 
properties of superclusters  of galaxies from the 
2dF Galaxy Redshift Survey (Einasto et al. \cite{e06a}, hereafter
Paper I; Einasto et al. \cite{ e06b}, Paper II; 
Einasto et al. (\cite{e07c}, Paper III). 

Already early studies of 
superclusters showed that the properties of galaxies depend on 
the environment where they reside in (Giovanelli, Haynes and 
Chincarini \cite{gio86}; Einasto and Einasto \cite{e87}). Recently, 
galaxy populations have been studied in some very rich superclusters 
(Haines et al. \cite{hai06} -- in the Shapley supercluster, Porter and 
Raychaudhury \cite{pr05} -- in the Pisces-Cetus supercluster). However, in 
these studies the properties of galaxies in rich clusters were studied 
in the core regions of superclusters only. These studies showed that 
rich clusters contain a large fraction of passive galaxies, while 
actively star forming galaxies are located between the clusters.

Our main goal in the present paper is the study of substructures and 
galaxy populations in the whole superclusters, not in the core regions 
only. Also we shall compare galaxy populations in different 
superclusters. We study the density distribution in superclusters.  As 
in the first paper, we shall calculate the Minkowski functionals, but now we 
apply this method for galaxies from different populations with the aim 
to compare substructures in superclusters as delineated by galaxies 
with these populations. This analysis shows whether we can use 
the Minkowski functionals in studies 
of galaxy populations. We compare galaxy populations in different 
superclusters -- the overall galaxy content and the distribution of galaxies 
of different luminosity, colour and spectral type in regions of 
different density. We also compare the richness and the galaxy 
content of groups in regions of different density.

The paper is composed as follows. In Section 2 we describe the galaxy 
data, the supercluster catalogue and the data on the richest 
superclusters. In Section 3 we study the distribution of environmental 
densities in superclusters, describe the use of the Minkowski functionals 
to study the morphology of superclusters, and present the results on 
supercluster morphology. In Section 4 we compare the properties of 
galaxies in rich and poor groups, in regions of different global density. 
We also compare the populations of galaxies in different superclusters. 
In the last sections we discuss our results and give the 
conclusions.

\section{Data}

\subsection{Rich supercluster data}

As in the previous paper, we used the 2dFGRS final release (Colless
\etal \cite{col01}; \cite{col03}), and the catalogue of superclusters
of galaxies from the 2dF survey (Paper I), applying a redshift limit
$z\leq 0.2$.  When calculating (comoving) distances we used a flat
cosmological model with the 
standard parameters, matter density
$\Omega_m = 0.3$, and dark energy density $\Omega_{\Lambda} = 0.7$
(both in units of the critical cosmological density).  Galaxies were
included in the 2dF GRS, if their corrected apparent magnitude 
$b_j$ lied in the interval from $b_1 = 13.5$ to $b_2 = 19.45$.  We
used weighted luminosities to calculate the luminosity density field 
on a grid with a cell size of 1~\Mpc\, smoothed with an Epanechnikov
kernel of the radius 8~\Mpc; this density field was used to find
superclusters of galaxies. We defined superclusters as connected
non-percolating systems with densities above a certain threshold
density; the actual threshold density used was 4.6 in units of the
mean luminosity density. A detailed description of the supercluster
finding algorithm can be found in Paper I.

In our analysis we also used the data about groups of galaxies from
the 2dFGRS (Tago et al. \cite{tago06}, hereafter T06).  Groups of
galaxies were determined using identical FoF parameters independently
of the global environment, fixing a certain local number density
threshold. Later we will use this division as a local density
indicator, to compare the properties of galaxies in various local and
global environments.  The catalogues of groups and isolated galaxies
can be found at \texttt{http://www.aai.ee/$\sim$maret/2dfgr.html}, the
catalogues of observed and model superclusters -- at
\texttt{http://www.aai.ee/$\sim$maret/2dfscl.html}.

For the present analysis we select three 
rich superclusters from the
catalogue of 2dF superclusters. Two of them are the richest superclusters in
our catalogue: the supercluster SCL126 in the Northern Sky, and the
supercluster SCL9 (the Sculptor supercluster) in the Southern Sky, according
to the catalogue by Einasto et al. (\cite{e2001}, hereafter E01). The third
supercluster, SCL10 (the Pisces-Cetus supercluster) in the Southern Sky is
relatively nearby. This supercluster was recently studied by Porter and
Raychaudhury (\cite{pr05}).

\begin{figure*}[ht]
\centering
\resizebox{0.32\textwidth}{!}{\includegraphics*{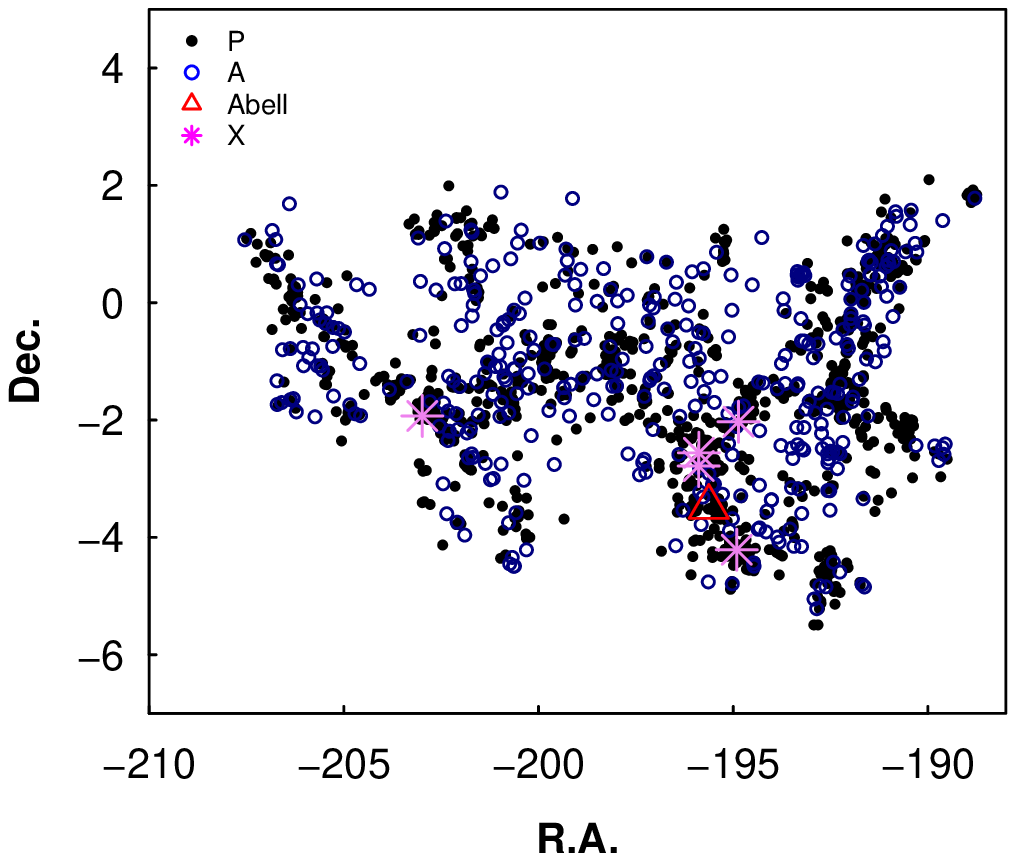}}
\resizebox{0.32\textwidth}{!}{\includegraphics*{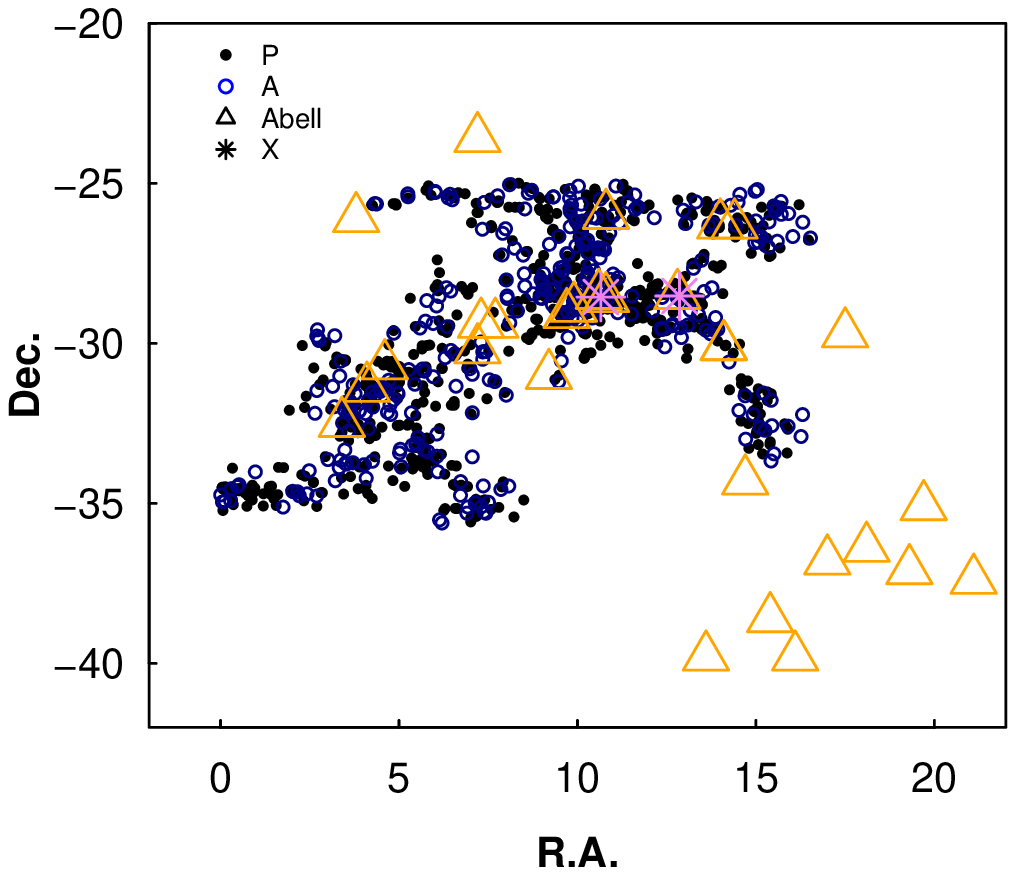}}
\resizebox{0.33\textwidth}{!}{\includegraphics*{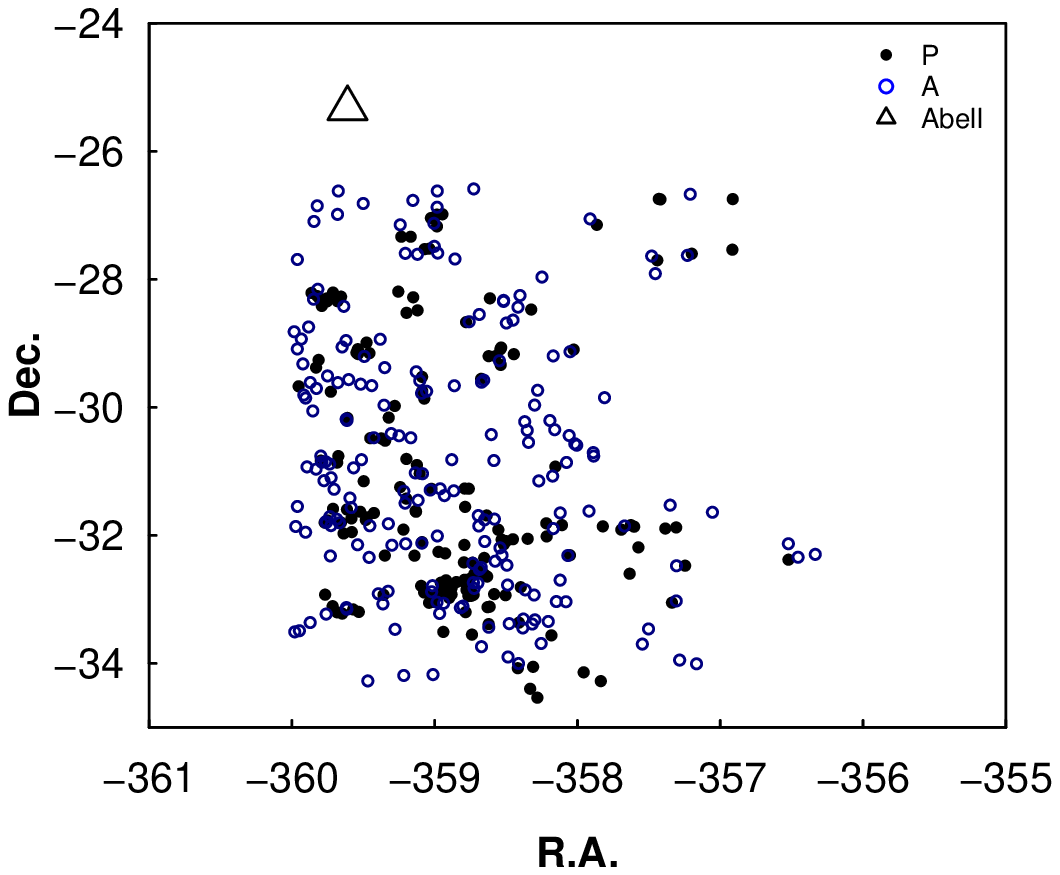}}
\caption{The sky distribution of galaxies, and Abell and X-ray clusters 
in superclusters SCL126, SCL9, and SCL10 (3 panels from left to right). 
Filled circles: passive (red) galaxies, empty circles: active (blue) galaxies,
triangles: Abell clusters, crosses: X-ray clusters.
}
\label{fig:radecx}
\end{figure*}

\begin{figure*}[ht]
\centering
\resizebox{0.32\textwidth}{!}{\includegraphics*{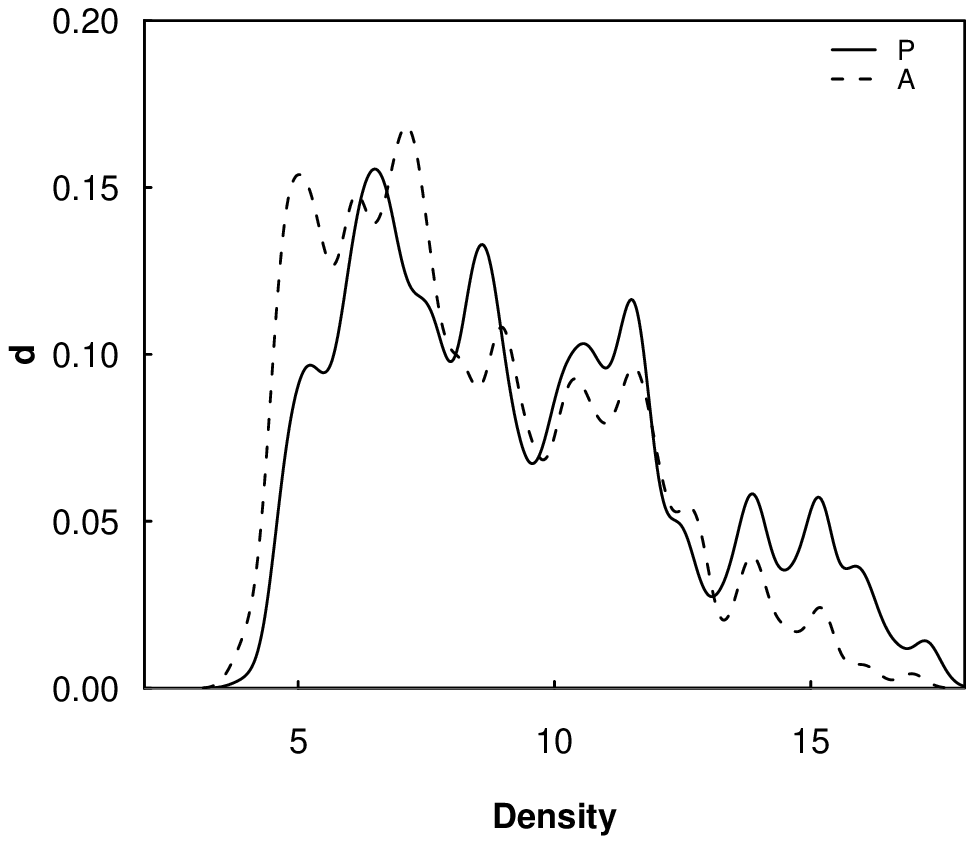}}
\resizebox{0.32\textwidth}{!}{\includegraphics*{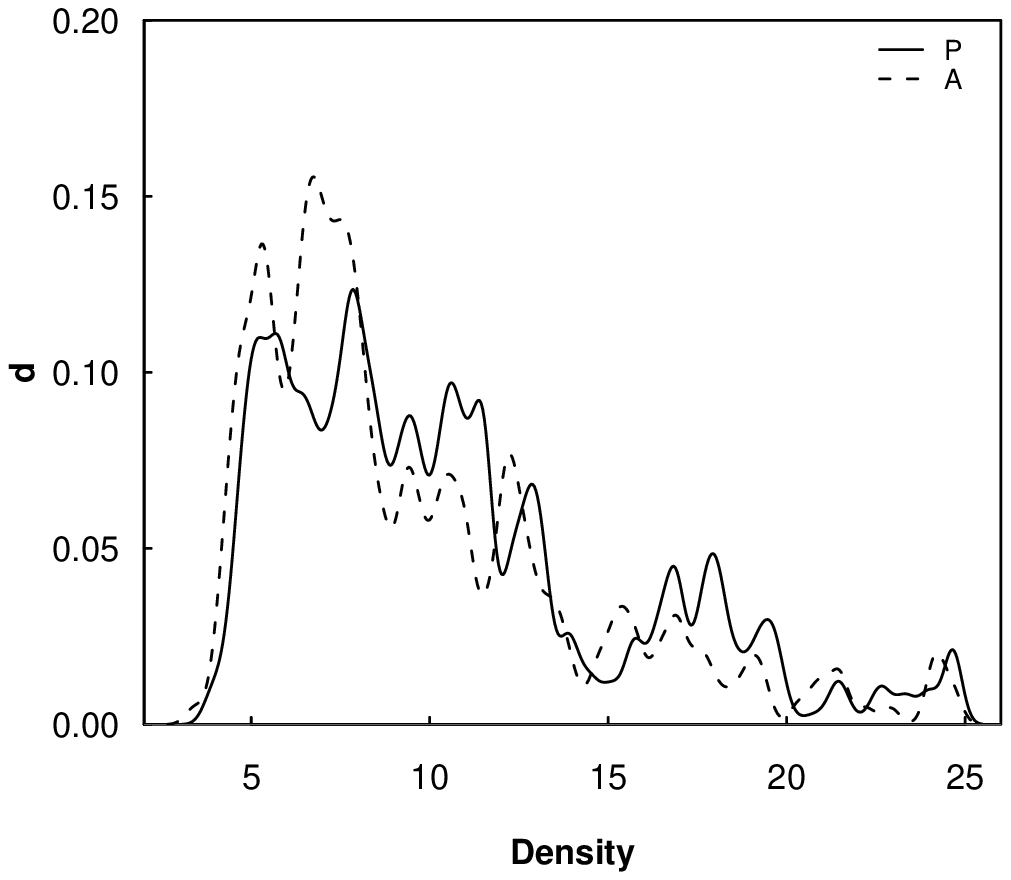}}
\resizebox{0.32\textwidth}{!}{\includegraphics*{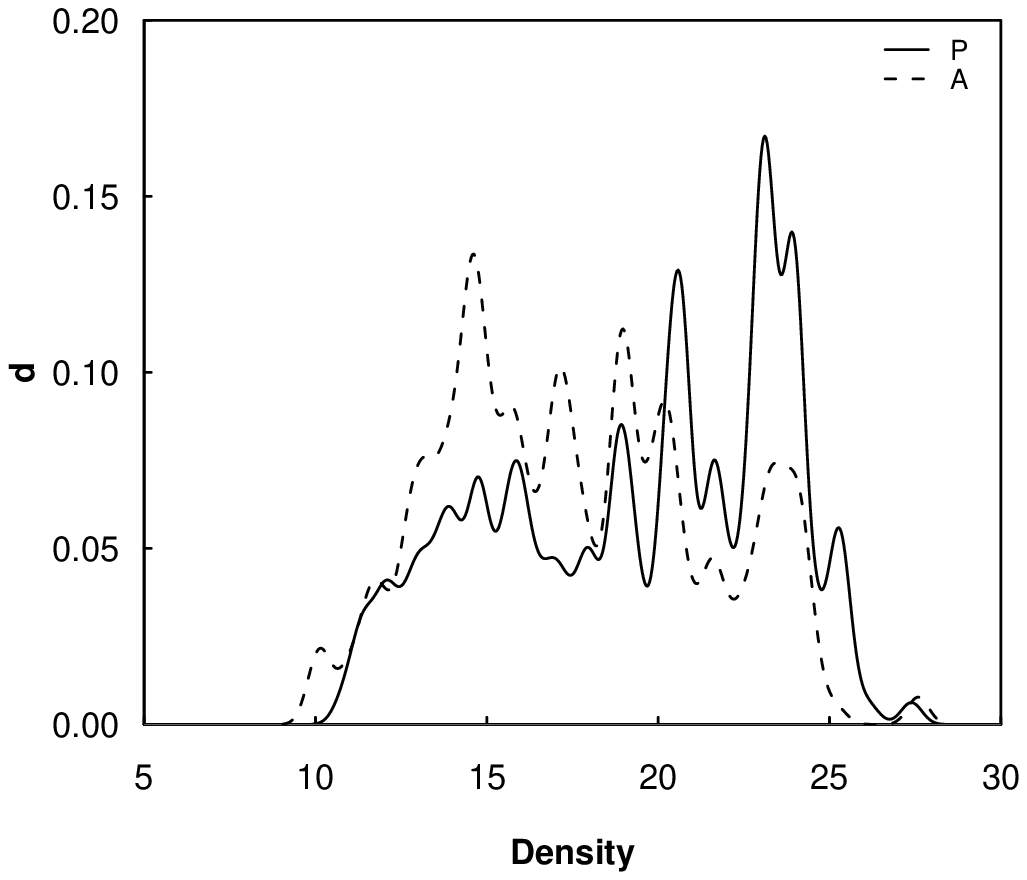}}
\caption{The distribution of luminosity densities in rich superclusters, at the
location of passive and actively star forming galaxies (defined by their 
colour index $col$).  
From left to right: SCL126,
SCL9, SCL10.  }
\label{fig:bfdensobs}
\end{figure*}

The general data on these superclusters are given in
Table~\ref{tab:1}.  In this Table we give the coordinates and
distances of superclusters, the numbers of galaxies, groups and Abell
and X-ray clusters in superclusters, the mean values of the luminosity
density field in superclusters and their total luminosities (from
Paper II).  In our morphological analysis we use volume-limited
samples of galaxies from these superclusters. The luminosity limits
for each supercluster sample are also given in Table~\ref{tab:1}. We
plot the distribution of galaxies, and Abell and X-ray clusters in
Fig.~\ref{fig:radecx}

The most prominent Abell supercluster in the Northern 2dF survey is the 
supercluster SCL126 (in E01, N152 in Paper I) that lies in the direction to the 
Virgo constellation. This supercluster has been also called the Sloan Great Wall 
(Vogeley et al. \cite{vogeley04}, Gott et al. \cite{gott05}, Nichol et al. 
\cite{nichol06}).  In E01 this supercluster contains 7 Abell clusters: A1620, 
A1650, A1651, A1658, A1663, A1692, A1750. Of these clusters, A1650, A1651, 
A1663, and A1750 are X-ray clusters.

The richest supercluster in the Southern Sky is the Sculptor
supercluster (SCL9 in E01, S34). This supercluster contains also
several X-ray clusters, and the largest number of Abell clusters in
our supercluster sample, 25. However, only 12 of them are located in
the region covered by the 2dF redshift survey.  These are: A88, A122,
A2751, A2759, A2778, A2780, A2794, A2798, A2801, and A2844, and two
X-ray clusters, A2811 and A2829.

Another nearby prominent supercluster in the Southern sky is the
Pisces-Cetus supercluster (SCL10 in E01, SCL S5), which contains the
rich X-ray cluster, Abell 2734. Only one of 19 Abell clusters from
this supercluster, A2683, is located within the 2dF survey boundaries.
This supercluster was recently described as a rich filament of Abell
clusters by Porter and Raychaudhury (\cite{pr05}).

\subsection{Properties of galaxies used in the present analysis}

To study the properties of galaxies in superclusters we used the
galaxy data as given in the 2dF redshift survey.  We characterize
galaxies by their luminosity, the spectral parameter $\eta$ and by the
colour index $col$ (Madgwick et al. \cite{ma02} and \cite{ma03a}, de
Propris et al. \cite{depr03}, Cole et al. \cite{cole05}) as follows
(see also Paper III).  In order to divide galaxies into populations of
bright and faint galaxies (B/F) we used a limit $B_{j}= -20.0$. 
In order to divide galaxies into populations of early and
late type galaxies (E/S), we used the spectral parameter limit 
$\eta = -1.4$, $\eta \leq -1.4$ for E, $\eta > -1.4$ for S. 
More detailed morphological types were defined as follows:
Type 1: $\eta < -1.4$; Type 2: $-1.4 \leq \eta < 1.1$; Type 3: $1.1
\leq \eta < 3.5$; Type 4: $3.5 \leq \eta$. Moreover, the spectral
parameter $\eta$ is correlated with the equivalent width of the
$H_{\alpha}$ emission line, thus being an indicator of the star
formation rate in galaxies. The value $\eta < 0.0$ corresponds to the
population of quiescent galaxies (q) and $\eta \geq 0.0$ to star-forming
galaxies (SFR). We also used information about colours of galaxies (the
rest-frame colour index, $col = (B - R)_0$ to divide galaxies into
populations of passive galaxies (P) and actively star forming galaxies (A);
for passive (red) galaxies $col \geq 1.07$.

\section{Substructure in rich superclusters}

\subsection{Density distribution}

We start our study of substructure in superclusters with the 
analysis of the environmental densities of galaxies $\delta$ in
superclusters (defined as the normalised value of the density field at
the location of galaxies).  This density field was calculated using the
Epanechnikov kernel of a radius 8~\Mpc\ (Sec. 2 and Paper I).

We plot the distribution of environmental densities for observed
superclusters in Fig.~\ref{fig:bfdensobs} for passive and actively
star forming galaxies (discriminated by the colour index $col$).

Fig.~\ref{fig:bfdensobs} shows that environmental densities in rich
superclusters are rather different: the maximum of densities in the
superclusters SCL9 and especially in SCL10 are much larger than in
SCL126.  The distribution of densities have several maxima and minima
-- this shows the presence of substructures with different
characteristic density in superclusters. The number of substructures
in SCL126 is smaller than in SCL9 -- an indication that the
supercluster SCL126 may be more homogeneous and less clumpy than SCL9.
In Sect. 3.2 we shall study this in more detail.

The distribution of environmental densities in the supercluster SCL10
shows, in accordance with  Fig.~\ref{fig:radecx}, that there is 
only one high density region in this supercluster (within the 2dF survey
region). There is an excess of passive galaxies in this clump or cluster.

Fig.~\ref{fig:bfdensobs} also shows how the maxima and minima in the
density distribution are traced by galaxies of different
properties. We see that although the overall density distributions for
passive and active galaxies are rather similar, there exist several
differences between these density distributions. At higher
environmental densities there is an excess of passive galaxies, at
lower densities -- 
an excess of active galaxies. In the supercluster
SCL126 the differences between the density distributions of passive
and active galaxies in both high density and the lowest density
regions (the core of the supercluster and the outer parts of the
supercluster, correspondingly) are larger than in other two
superclusters. This is an indication of large-scale morphological
segregation in superclusters, which in SCL126 is stronger than in SCL9
and SCL10. Such a segregation can be also found, if we study
environmental densities around red and blue galaxies (classified by
the spectral parameter $\eta$); red galaxies have higher densities
around them.

For the analysis of galaxy populations in regions of different density in
superclusters we divide superclusters into three regions of different
density, according to the location of minima in the density
distribution, which separate regions of different environmental
densities. The density limits for each region are given in
Table~\ref{tab:5}.

In the supercluster SCL126 the regions of highest density form two
main spatially separate regions. One of them is the main core of the
supercluster, containing four Abell clusters (three of them are also
X-ray sources).  This region has a diameter of about 10~\Mpc\ (Einasto
et al.  \cite{e03d}).  One X-ray cluster in this supercluster, Abell
1750, is located in another region of high density. This cluster is a
merging binary cluster (Donelly et al. \cite{don}; Belsole et
al. \cite{bel}).

In the supercluster SCL9 the regions of highest density form three
main separate concentrations of galaxies. One of them, the main center
of SCL9, contains 5 Abell clusters and one X-ray cluster. There are
Abell clusters also in two other regions of highest density.  However,
some Abell clusters are located in regions of lower density, $D2$, and
one X-ray cluster is located in a region of relatively low density,
$D3$. In this supercluster the total number of Abell clusters is
larger than in the supercluster SCL126, but they do not form such a
high concentration of Abell and X-ray clusters that is observed in the
core region of SCL126. Most of X-ray clusters in SCL9 are located
outside the region covered by the 2dF redshift survey.

In the supercluster SCL10 most Abell clusters remain outside the region 
covered by the 2dF survey (see also Porter and Raychaudhury 
\cite{pr05}). The environmental densities of galaxies in this 
supercluster are rather high. 

In general the properties of individual 
superclusters, including the distribution of environmental densities, 
vary strongly.

\subsection{Minkowski Functionals } 

Next we study the substructure in the distribution of galaxies from
different populations in superclusters using Minkowski functionals  
This study is of exploratory nature since this is the
first time when these functionals are used for studies of galaxy
populations. Using Minkowski functionals, we can see in detail how the
morphology of superclusters is traced by galaxies of different type.

\begin{figure*}[ht]
\centering
\resizebox{0.30\textwidth}{!}{\includegraphics*{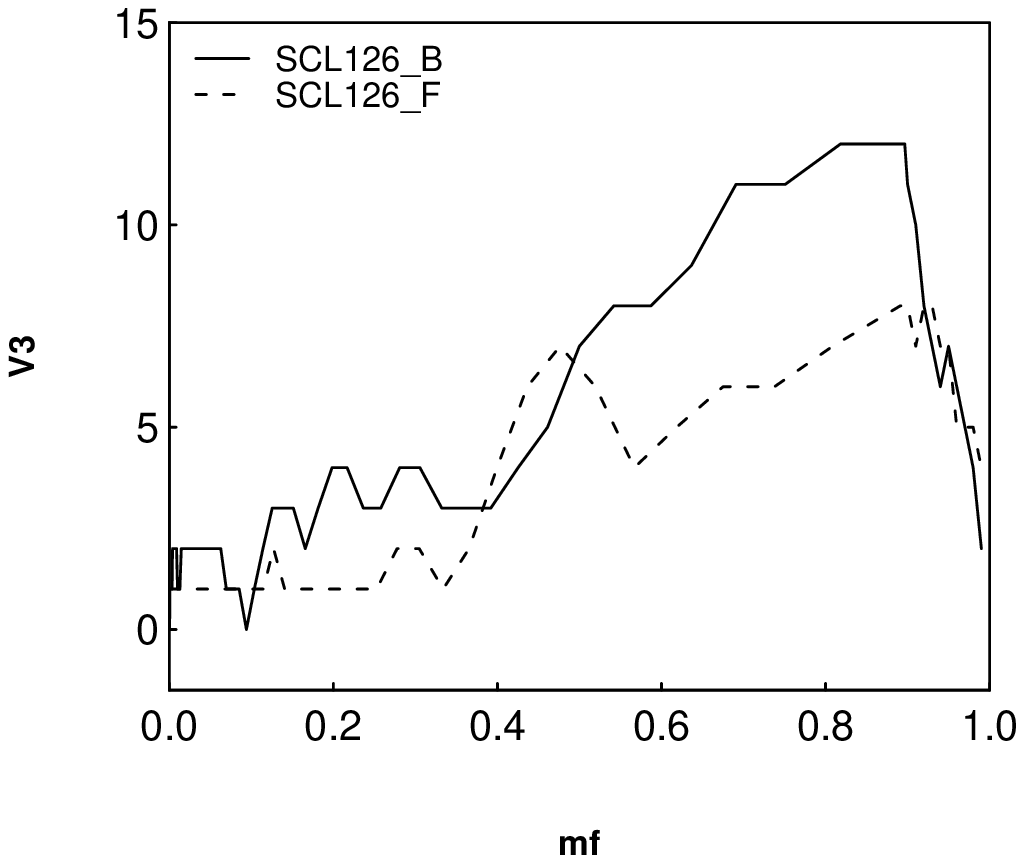}}
\resizebox{0.30\textwidth}{!}{\includegraphics*{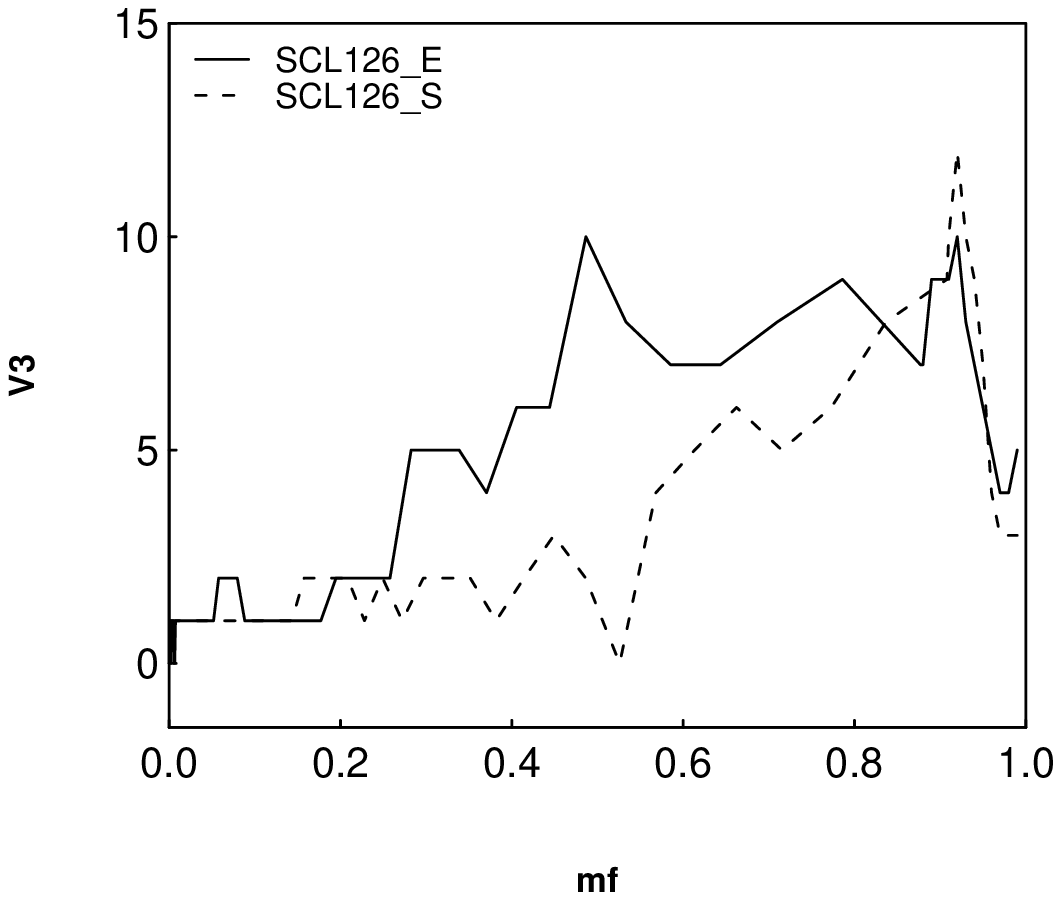}}
\resizebox{0.30\textwidth}{!}{\includegraphics*{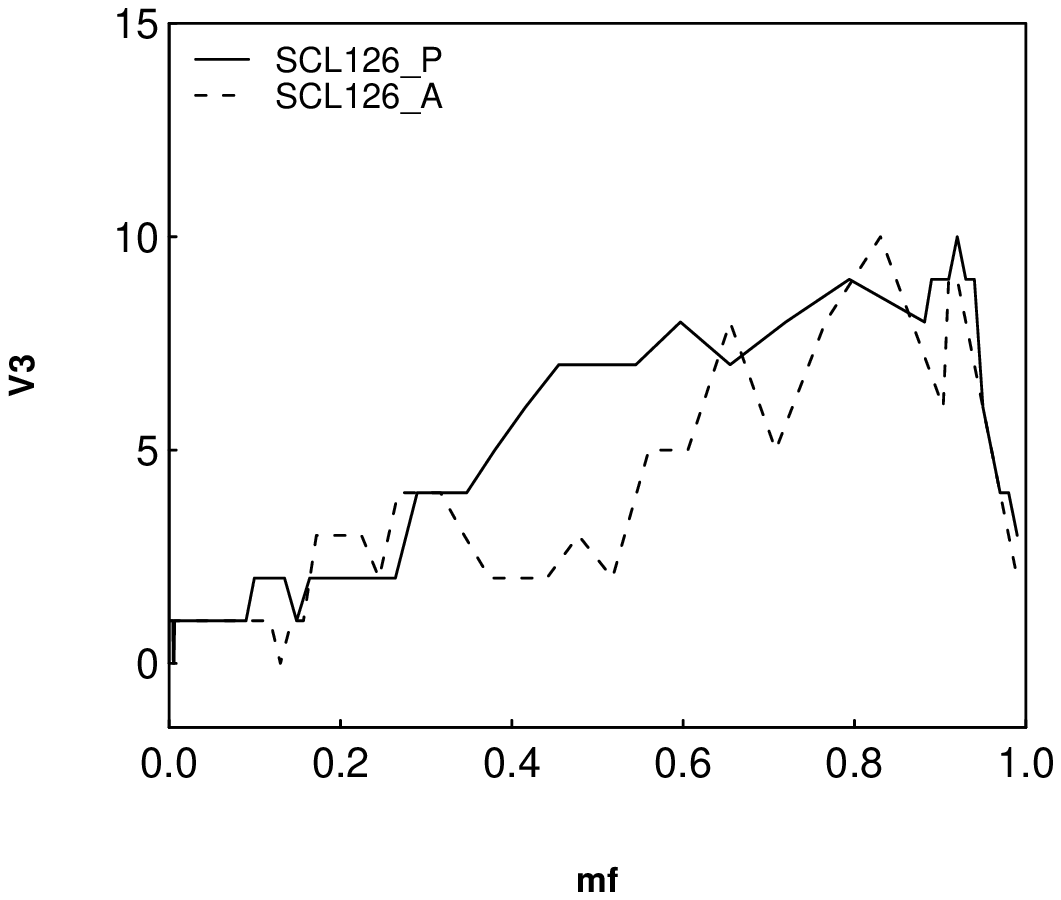}}
\hspace*{2mm}\\
\resizebox{0.30\textwidth}{!}{\includegraphics*{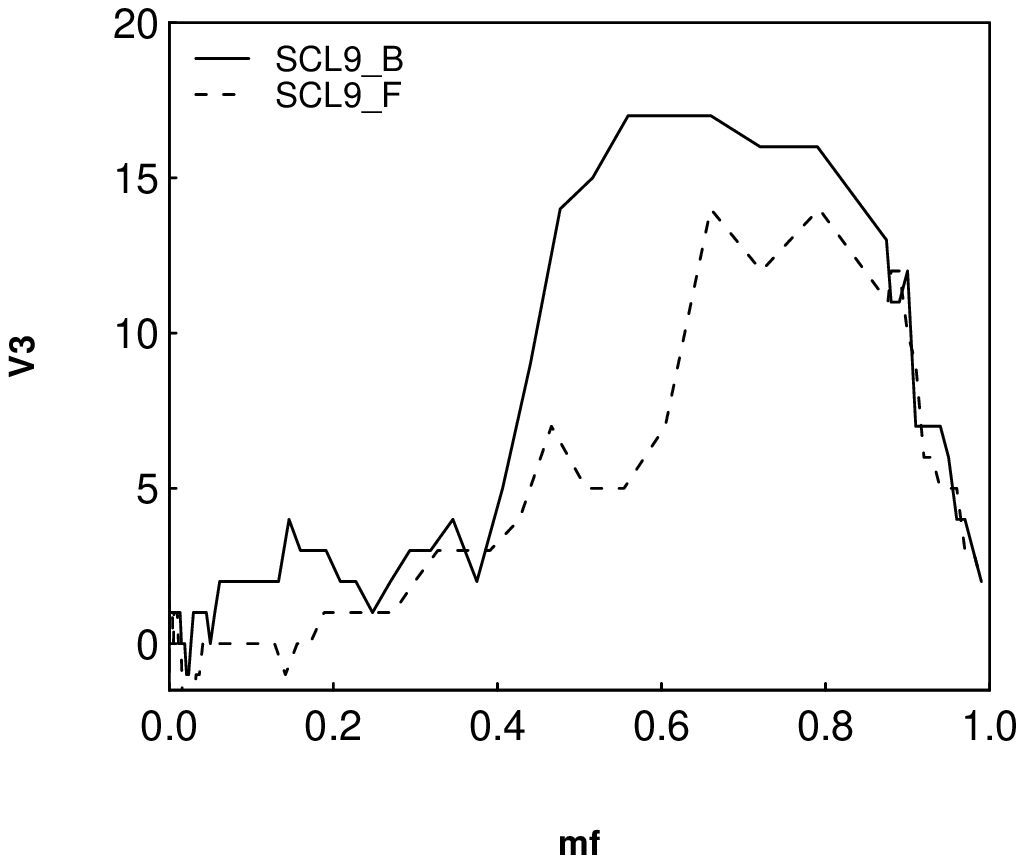}}
\resizebox{0.30\textwidth}{!}{\includegraphics*{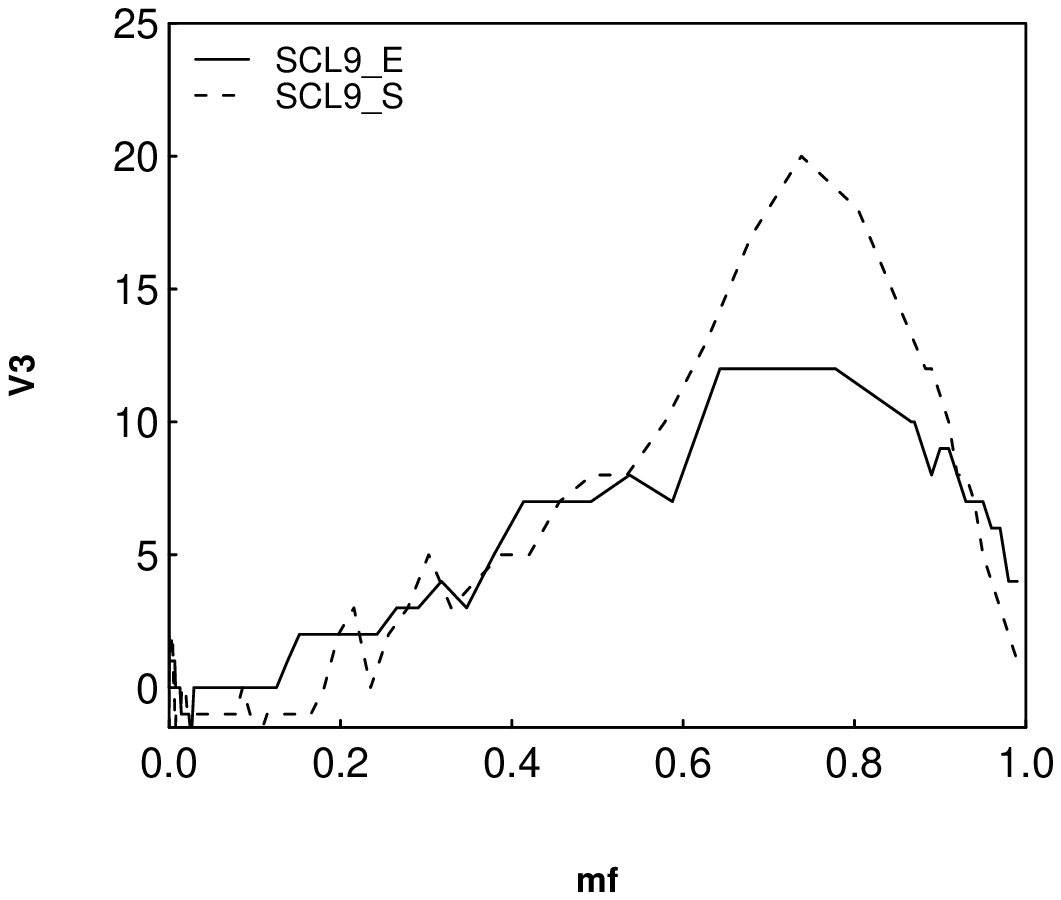}}
\resizebox{0.30\textwidth}{!}{\includegraphics*{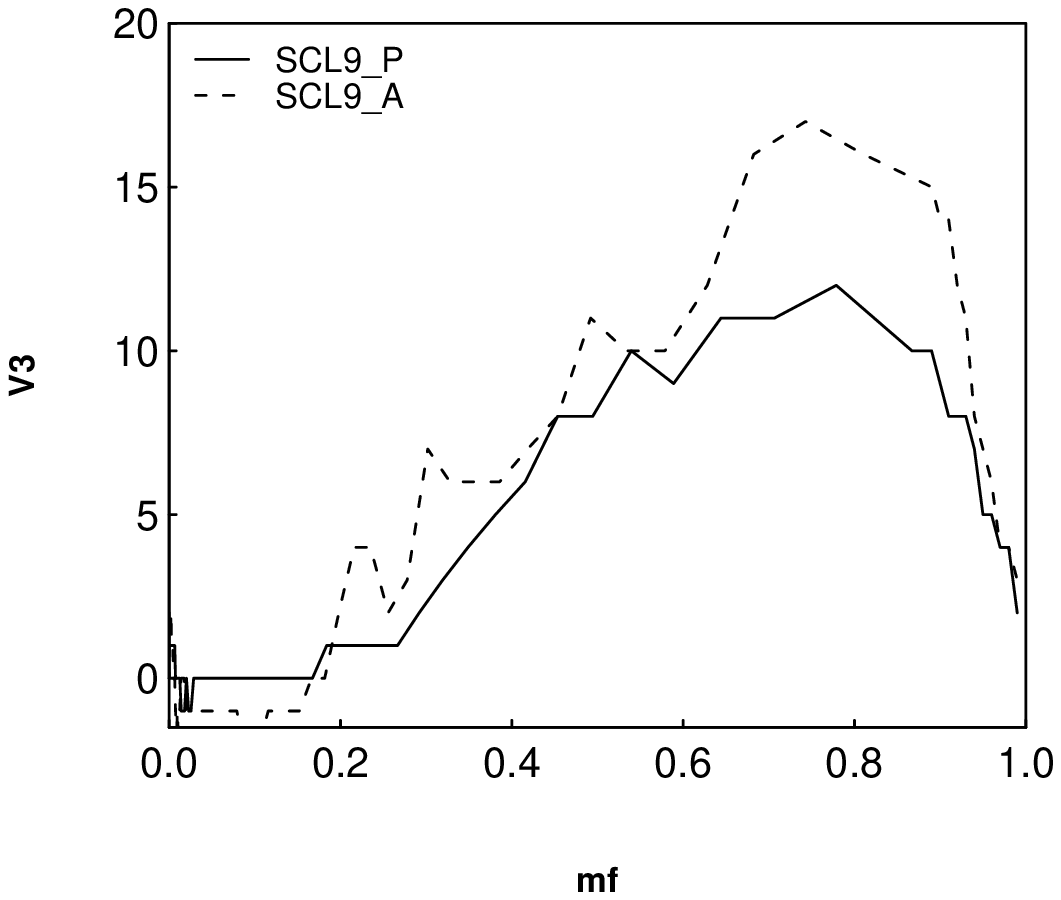}}
\hspace*{2mm}\\
\resizebox{0.30\textwidth}{!}{\includegraphics*{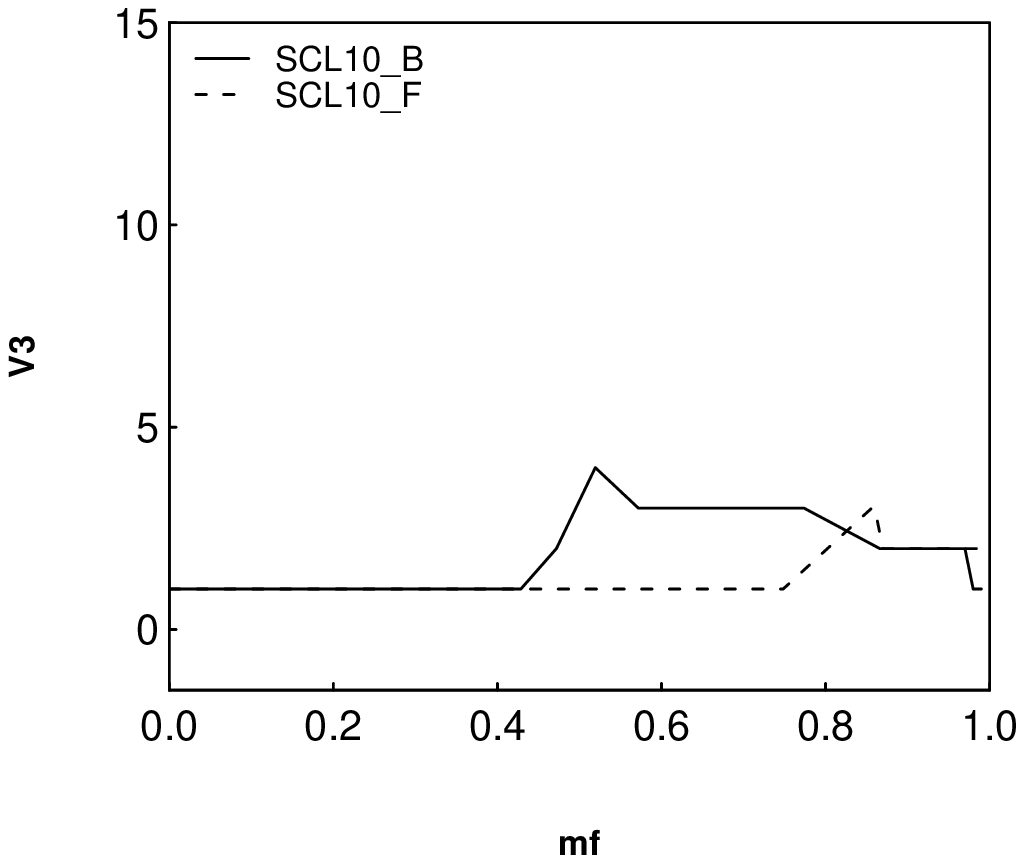}}
\resizebox{0.30\textwidth}{!}{\includegraphics*{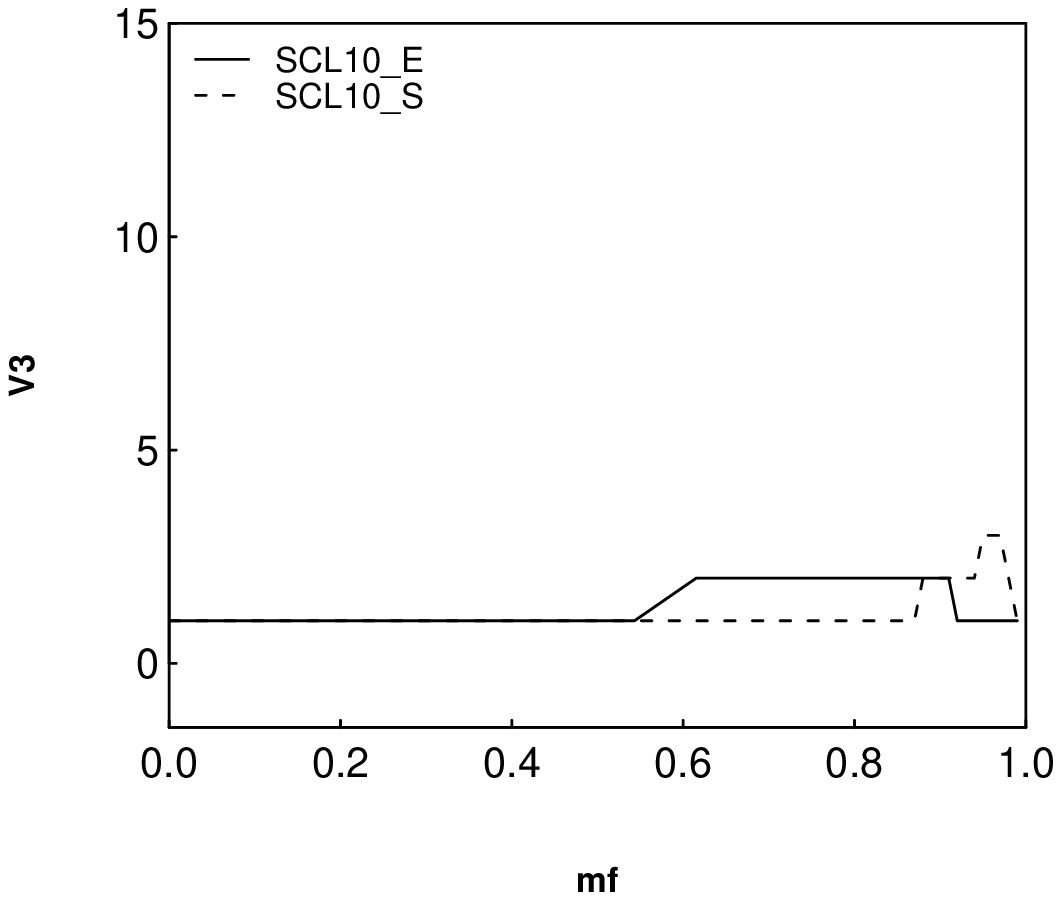}}
\resizebox{0.30\textwidth}{!}{\includegraphics*{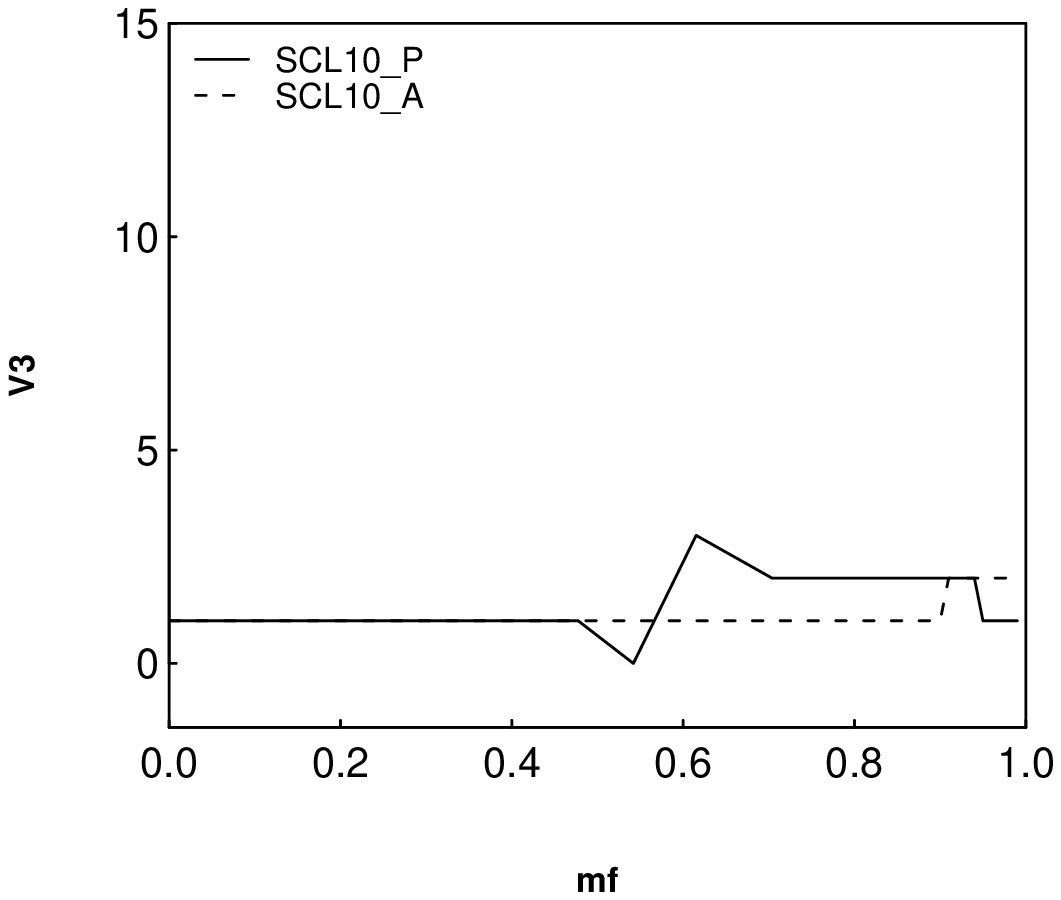}}
\caption{The Minkowski functional $V_3$ (the Euler Characteristic)
vs the mass fraction $m_f$ for the bright (B, $M \leq -20.0$) and 
faint (F, $M > -20.0$) galaxies (left panel), for early and late 
type galaxies (middle panels) and for passive and active galaxies
(right panels). Upper panels: the supercluster SCL126, 
middle panels: the supercluster SCL9, and lower panels: 
the supercluster SCL10.  
}
\label{fig:v3}
\end{figure*}

The supercluster geometry (morphology) is given by their outer
(limiting) isodensity surface, and its enclosed volume. When
increasing the density level over the threshold overdensity
$\delta=4.6$ (sect. 2.1), the isodensity surfaces move into the
central parts of the supercluster.  The morphology and topology of the
isodensity contours is (in the sense of global geometry) completely
characterized by the four Minkowski functionals $V_0$ -- $V_3$.

For a given surface the four Minkowski functionals are as follows:
\begin{itemize}
\item The first, $V_0$, is the enclosed volume $V$. 
\item The second, $V_1$, is proportional to the area of the surface 
$S$, $V_1 = {1\over6} S $.
\item 
The third, $V_2$, is proportional to the integrated mean curvature $C$, 
\[V_2 = \frac1{3\pi} C, \qquad 
C=\frac12\int_S\left(\frac1{R_1}+\frac1{R_2}\right)\,dS,
\]
where $R_1$ and $R_2$ are the two local principal radii of curvature.
\item 
The fourth, $V_3$, is proportional to the integrated Gaussian 
curvature (or Euler characteristic) $\chi$, 
\[V_3=\frac12\chi,\qquad
  \chi = \frac1{2\pi}\int_S\left(\frac1{R_1R_2}\right)dS.
\]
\end{itemize}

 At high (low) densities this functional gives us the number of isolated 
  clumps (voids) in the sample (Martinez et al. 2005; 
  Saar et al. \cite{saar06}).
 
To obtain the density field for estimating the Minkowski
functionals, we used a kernel estimator with a $B_3$ box spline as the
smoothing kernel, with the total extent of 16~\Mpc\ (for a detailed
description see Saar et al. \cite{saar06} and Paper RI).  For the
argument labeling the isodensity surfaces, we chose the mass fraction
$m_f$ -- the ratio of the mass in regions with density lower than the
density on the surface, to the total mass of the supercluster. When
this ratio runs from 0 to 1, the isosurfaces move from the outside
into the center of the supercluster.

  In Paper RI we calculated the Minkowski functionals for full
 superclusters and showed that the fourth Minkowski functional $V_3$
 describes well the clumpiness of superclusters.  The region of the
 highest density, $D1$, approximately corresponds to the region where
 the Minkowski functional $V_3$ has a maximum (see Paper RI for
 details).  The morphology of superclusters is characteristic to
 multibranching filaments.

Now we shall find the Minkowski functional $V_3$ for rich superclusters
separately for galaxies from different populations, as marked by their
luminosity, the spectral parameter $\eta$ and the colour index $col$
(Fig.~\ref{fig:v3}).

At small mass fractions the isosurface includes the whole
supercluster.  As we move to higher mass fractions, the isosurfaces
include only the higher density parts of superclusters. Individual
high density regions in a supercluster, which at low mass fractions
are joined together into one system, begin to separate from each
other, 
and the value of the fourth Minkowski functional ($V_3$)
increases. At a certain density contrast (mass fraction) $V_3$ has a
maximum, showing the largest number of isolated clumps in a given
supercluster. At still higher density contrasts only 
the high
density peaks contribute to the supercluster and the value of $V_3$
decreases again.

Fig.~\ref{fig:v3} (upper left panel) shows the $V_3$ curves for bright 
and faint galaxies in the supercluster SCL126. At small values of the 
mass fraction $m_f$ the values of $V_3$ are small. At a mass fraction 
$m_f \approx 0.4 $ the values of $V_3$ of both bright and faint galaxies 
increase rapidly; for bright galaxies it reaches the value of about 10, 
while for faint galaxies the value of $V_3$ remains about 5. For the 
mass fraction interval of about 0.4--0.9 the values of $V_3$ remain 
almost unchanged. This indicates that the overall morphology of the 
supercluster SCL126 is rather homogeneous, which is characteristic to a 
rich filament with several branches (see Paper RI for details). In this 
supercluster, the distribution of bright galaxies is more clumpy than 
the distribution of faint galaxies. In other words, bright galaxies are 
located in rich groups while fainter galaxies form a less clumpy 
population around them (a population of poor groups and of galaxies not 
belonging to groups; we shall discuss their distribution in Sect. 4). 
The peaks of $V_3$ values at very high mass fractions are due to high-
density cores in this supercluster.

The values of the fourth Minkowski functional, $V_3$, for galaxies of
different type (Fig.~\ref{fig:v3}, upper middle panel) show that
the clumpiness of early type galaxies starts to increase at lower values
of the mass fraction $m_f$.  During the mass fraction interval
0.4--0.9 this value changes only a little, and has a peak value at the
mass fraction of about 0.9.  The distribution of late-type galaxies in
this supercluster is much more homogeneous (less clumpy), as show the
values of $V_3$. The number of isolated clumps of these galaxies grows
only at rather high mass fraction values, $m_f > 0.6$, and has a peak
at $m_f \approx 0.9$.

The behaviour of the fourth Minkowski functional for passive (red)
galaxies is rather similar to that of early type galaxies. In the case
of actively star forming (blue) galaxies we see that these galaxies
form some isolated clumps of star forming galaxies already at
relatively low mass fraction values, at higher mass fractions some of
them do not to contribute to the supercluster any more; the value of
$V_3$ decreases and then increases again. At very high values of the
mass fraction, $m_f > 0.9$, these galaxies again do not to contribute
to the supercluster, and the value of $V_3$ decreases rapidly. The
value of $V_3$ for passive galaxies is larger than for active galaxies
at almost all density levels (mass fraction values). Passive galaxies
are typically more strongly clustered than active galaxies, thus their
larger clumpiness is expected.

Next let us study the values of $V_3$ for galaxies from different 
populations in the supercluster SCL9 (Fig.~\ref{fig:v3}, middle panels). 
Here we see that the $V_3(m_f)$ curve is rather different from that for 
SCL126: at small mass fractions, the values of $V_3$ for bright and 
faint galaxies (left panel) are small, but at a mass fraction value of 
about 0.4, the values of $V_3$ increase. This increase is more rapid for 
bright galaxies than for faint galaxies. The values of $V_3$ reach 
maximum at a mass fraction value of about 0.6, the maximum values of 
$V_3$ are larger than in the supercluster SCL126. This indicates that 
the overall morphology of the supercluster SCL9 is more clumpy than that 
of the supercluster SCL126.

In Paper RI we showed that according to Minkowski functionals, the
supercluster SCL9 can be described as a multispider rather than a rich
filament, i.e. this supercluster consists of a large number of
relatively isolated clumps or cores connected by relatively thin
filaments, in which the density of galaxies is too low to contribute
to the higher density parts of this supercluster.  For
the $V_3(m_f)$ curve this means that the value of $V_3$ increases,
reaches a maximum and decreases again.

In the middle panel of Fig.~\ref{fig:v3} ($V_3$ for early and late
type galaxies) we see, in consistency with the above picture, a
continuous increase of the number of isolated clumps as delineated by
galaxies of different type, and then a rapid decrease of the number of
the clumps.  Surprisingly, the distribution of late type galaxies is
even more clumpy than the distribution of early type galaxies: at all
density levels the value of $V_3$ for early type galaxies (and also
for passive galaxies, right panel) is smaller than for late type
galaxies (and for active galaxies). This indicates that late type,
star-forming galaxies are located in numerous clumps around early
type, passive, red galaxies.

Thus what we see here is that differences in the distribution of 
galaxies from different populations in individual superclusters are
related 
to a different overall morphology of these superclusters.

The $V_3$ values for different galaxy populations in supercluster SCL10 
(Fig.~\ref{fig:v3}, lower panels) are consistent
with the expectations of a supercluster with
one core region (we remind that a bigger part of this supercluster
remains outside of our sample boundaries).

\subsection{The richness of groups in regions of different density} 

\begin{figure*}[ht]
\centering
\resizebox{0.26\textwidth}{!}{\includegraphics*{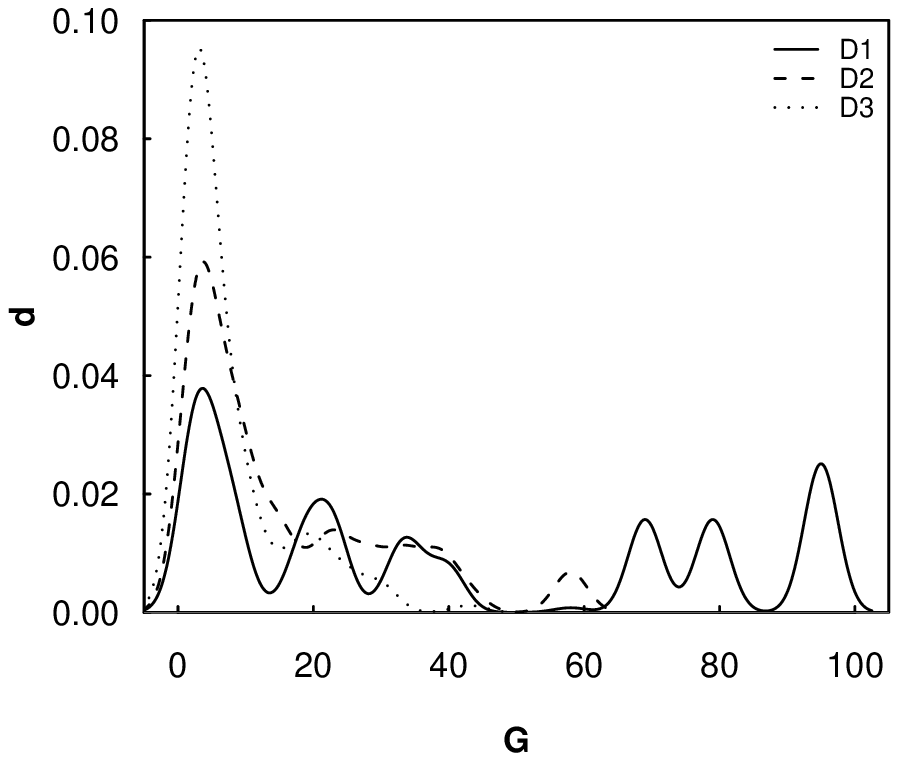}}
\resizebox{0.27\textwidth}{!}{\includegraphics*{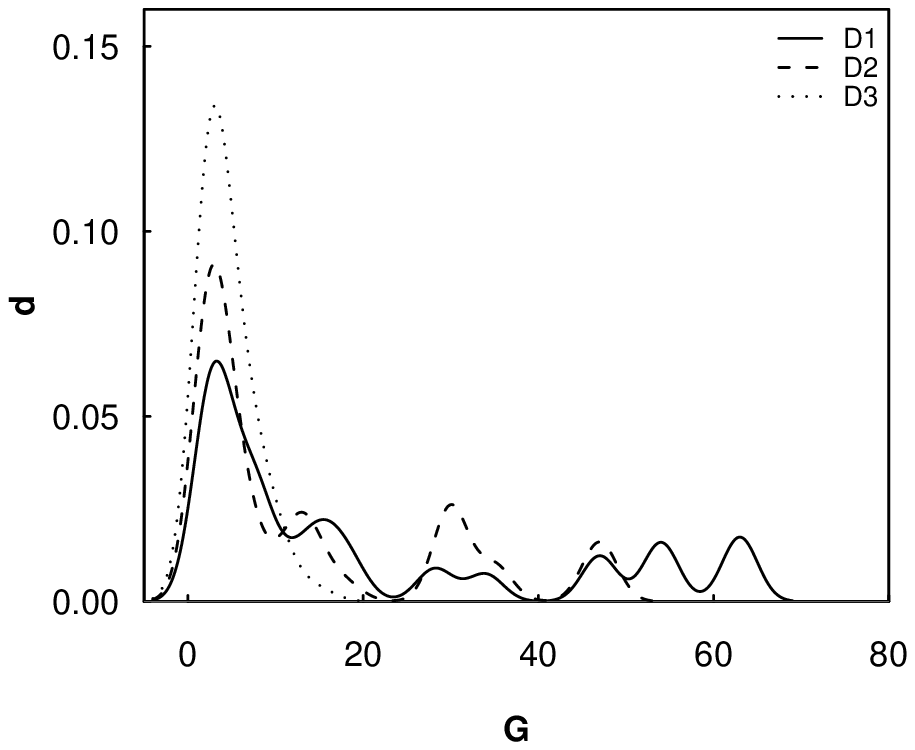}}
\resizebox{0.27\textwidth}{!}{\includegraphics*{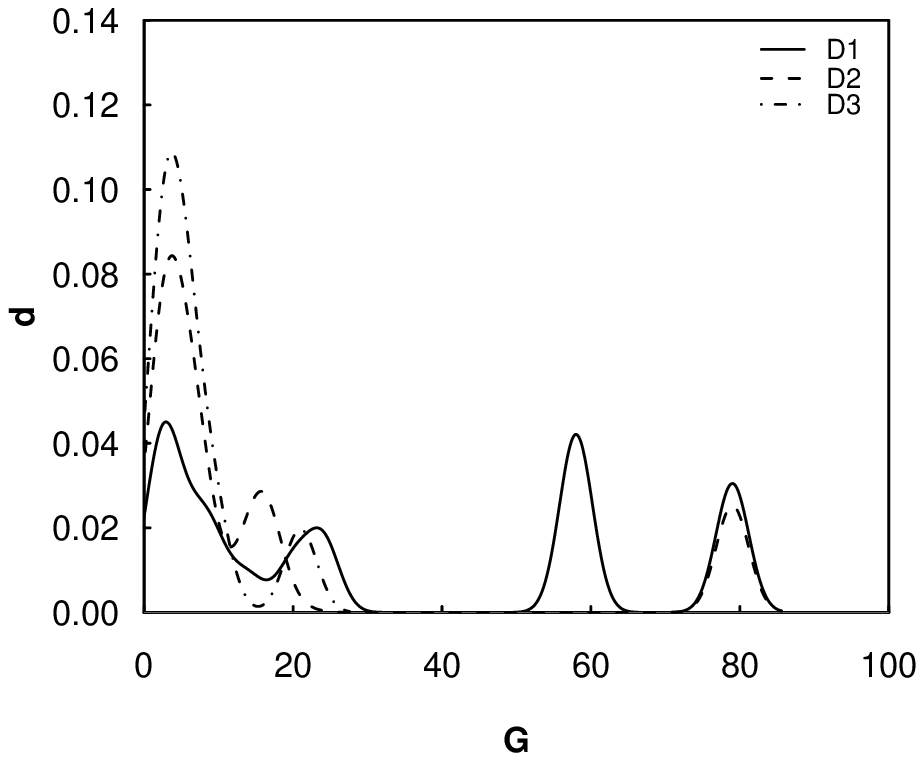}}
\caption{The number of galaxies in groups, $G$, 
in regions of different density in rich superclusters. 
From left to right: SCL126, SCL9,  SCL10.
}
\label{fig:grd14}
\end{figure*}

Groups of galaxies are additional indicators of substructure in
superclusters. We divide groups by their richness as follows: rich
($N_{gal} \geq 10$) groups and clusters (we denote this sample as
$Gr_{10}$) and poor groups ($N_{gal} < 10$, $Gr_2$). In
Table~\ref{tab:5} we give the fractions of galaxies in these
groups. To study where the groups of different richness reside inside
superclusters we plot the richness of groups, $G$, (see T06) in regions of
different density in superclusters (Fig.~\ref{fig:grd14}).  This
figure shows that in regions of high density one can find groups of
all richness; the richest groups in superclusters are located mostly
in these regions. There are no very rich groups in regions of lower
densities, 
and the richest groups in lower density regions are relatively poor.

{\scriptsize
\begin{table*}[ht]
\caption{Properties of galaxies in rich and poor groups, and of
isolated galaxies, in regions of different density in rich superclusters. }
\tiny
\begin{tabular}{llrclrclrclrcl} 
\hline 
         &     &      & SCL126 &       &       &   SCL9 &      &     &   &  SCL10 &     \\      
\hline 
(1)&(2)&(3)&(4)&(5)&(6)&(7)&(8)&(9)&(10)&(11)&(12)&(13)&(14)\\
\hline 
Region   & $All$ &  $D1$ &  $D2$  &  $D3$ &-19.50&$All$&  $D1$ &  $D2$ & $D3$ &$All$& $D1$ &  $D2$  & $D3$\\
D$_{lim}$&       &  12.0 &  8.0   &       &   &   &  12.0 &  9.0  &      &   & 21.0 &  16.0  &     \\
N$_{gal}$&       &  234  &  501   &   573 &   &   &  342  &  264  & 570  &   & 239  &  271   &  247\\
F$gr_2$  &       &  0.29 &  0.41  &  0.53 &   &   &  0.40 & 0.41  & 0.59 &   & 0.24 &  0.19  & 0.43\\
F$gr_{10}$&      &  0.57 &  0.34  &  0.17 &   &   &  0.40 & 0.29  & 0.05 &   & 0.54 &  0.30  & 0.11\\
$E/S$ &&&&&&&&&&&&&\\                                        
~~$All$    & 1.65   &2.77   & 1.81 &  1.28  &1.87& 1.96& 2.35 & 2.81  & 1.54 &0.84& 1.22 &  0.83  & 0.59\\
~~$Gr_{10}$& 3.54   &  4.54 & 3.28 &  3.00  &4.02& 3.65& 4.07 & 5.42  & 1.15 &1.75& 1.76 & 1.87 &  1.45 \\
~~$Gr_2$   & 1.48   &  1.87 & 1.61 &  1.33  &1.74& 2.03& 2.05 & 2.72  & 1.85 &0.71& 1.00 & 0.73 &  0.58 \\
~~$IG$     & 0.89   &  1.21 & 1.08 &  0.71  &0.91& 1.30& 1.11 & 1.79  & 1.20 &0.45& 0.50 & 0.44 &  0.44 \\
$q/SF$ &&&&&&&&&&&&&\\                                         
~~$All$    & 3.43   &  4.82 & 3.69 &  2.87  &4.25 & 4.29& 5.26 & 6.11&3.37 &1.60& 3.06&  1.73 &  1.09 \\
~~$Gr_{10}$& 7.98   &  8.50 & 7.55 &  8.09  &10.14& 7.64&10.42&10.01& 2.11 &3.58& 4.29 & 3.13 &  2.37 \\
~~$Gr_2$   & 3.11   &  3.93 & 3.19 &  2.92  &3.78& 4.24&4.83& 5.75 & 3.07  &1.33& 2.72 & 1.28 &  1.06 \\
~~$IG$     & 1.95   &  1.59 & 2.29 &  1.81  &2.35& 3.23&2.53& 4.57 & 3.12  &0.98& 2.00 & 0.76 &  0.88 \\
$P/A$ &&&&&&&&&&&&&\\                                         
~~$All$    & 2.53   &  3.98 & 2.85 &  1.95  &2.77& 2.45& 2.76 &  3.63 & 1.95 &1.25& 2.46 &  1.05 & 0.83 \\
~~$Gr_{10}$& 5.44   &  7.37 & 6.17 &  3.35  &6.21& 4.46& 5.00 &  6.00 & 1.55 &3.55& 4.65 & 2.67 &  2.38 \\
~~$Gr_2$   & 2.39   &  2.63 & 2.69 &  2.18  &2.58& 2.54& 2.34 &  3.36 & 2.42 &0.98& 1.32 & 0.99 &  0.83 \\
~~$IG$     & 1.35   &  1.58 & 1.47 &  1.23  &1.43& 1.67& 1.48 &  2.71 & 1.46 &0.68& 1.47 & 0.52 &  0.59 \\
$B/F$ &&&&&&&&&&&&&\\                                      
~~$All$    & 0.44   &  0.38 & 0.44 &  0.46  &0.75& 0.90& 0.98 & 0.76  & 0.93 &0.07& 0.09 & 0.04 &  0.09 \\
~~$Gr_{10}$& 0.52   &  0.48 & 0.50 &  0.61  &0.85& 0.96& 1.00 & 0.97  & 0.75 &0.06& 0.06 & 0.06 &  0.08 \\
~~$Gr_2$   & 0.41   &  0.32 & 0.45 &  0.49  &0.71& 0.88& 1.01 & 0.82  & 0.99 &0.07& 0.16 & 0.04 &  0.11 \\
~~$IG$     & 0.33   &  0.14 & 0.35 &  0.34  &0.56& 0.76& 0.81 & 0.53  & 0.84 &0.04& 0.06 & 0.01 &  0.05 \\
\label{tab:5}                        
\end{tabular}
{\it  }

The columns in the Table are as follows:
\noindent column 1: Population ID and number ratio type. 
D$_{lim}$ -- the density limit, N$_{gal}$ -- the number of galaxies in a given
density region, F$gr_2$ -- the fraction of galaxies in poor groups
with 2--9 member galaxies, F$gr_{10}$ -- the fraction of galaxies
in rich groups with at least 10 member galaxies. $E/S$ -- 
the ratio of the numbers of early and late type galaxies,
$q/SF$ the number of quiesent and star-forming galaxies, 
$P/A$ -- the ratio of the numbers of passive and active galaxies, 
$B/F$ -- the ratio of the numbers of bright 
($M_{bj}\leq -20.0$) and faint ($M_{bj} > -20.0$) galaxies.
The subheaders are: $All$ -- all galaxies, $Gr_{10}$ -- rich groups,
$Gr_2$ -- poor groups, $IG$ -- isolated galaxies.

\noindent Columns 2--6: the ratio of the numbers of galaxies
in given populations in the supercluster SCL126. $All$ -- all galaxies,
$D1$--$D3$ -- galaxies in these density regions,
$-19.50$ -- all galaxies brighter than $M_{bj}\leq -19.5$.
Columns 7--10 and 11--14: the same ratios 
in similar populations in the superclusters SCL9 and SCL10
(without the column $-19.50$).
\end{table*}
}

\section{Populations  of galaxies  in rich 
  superclusters}

\begin{figure*}[ht]
\centering
\resizebox{0.28\textwidth}{!}{\includegraphics*{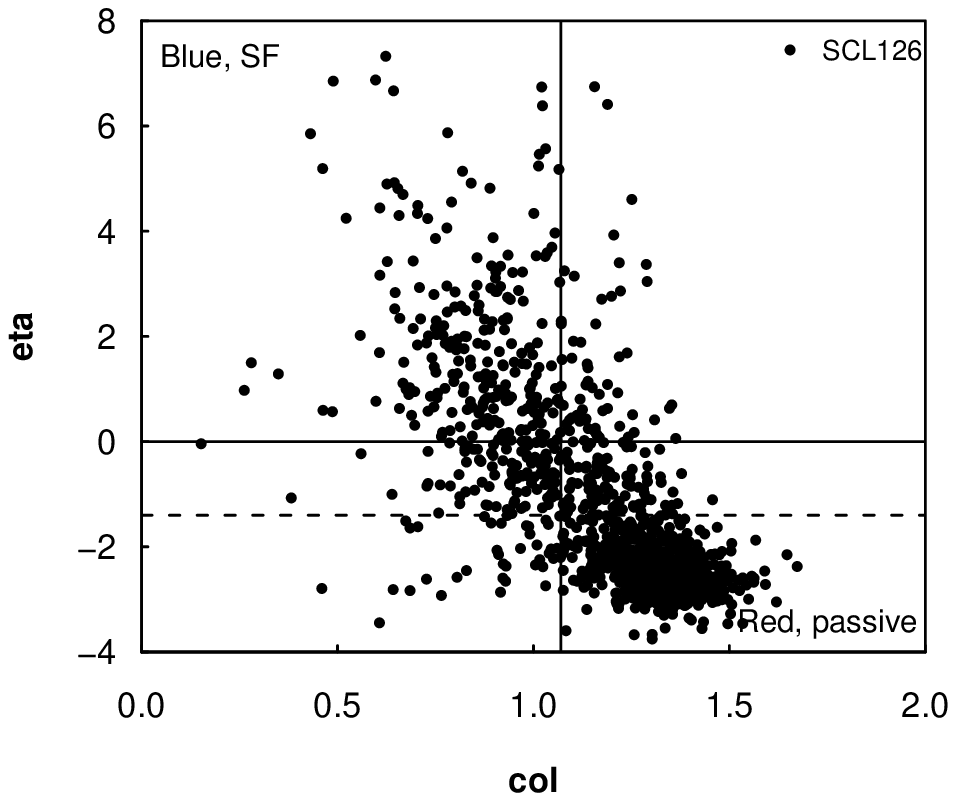}}
\resizebox{0.28\textwidth}{!}{\includegraphics*{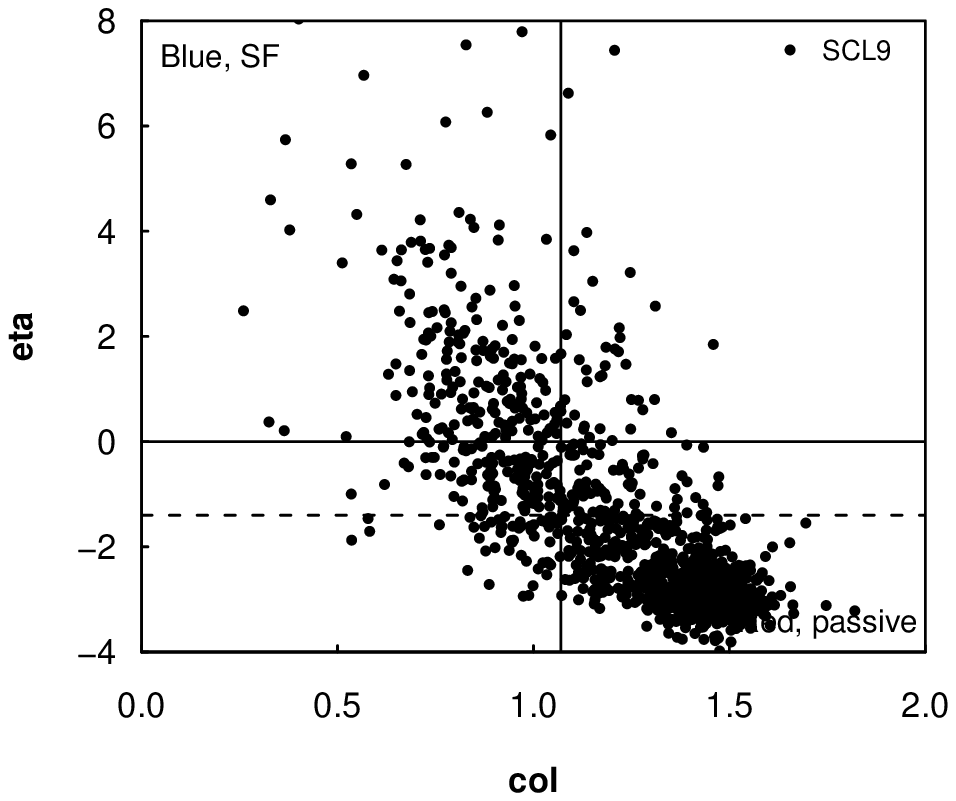}}
\resizebox{0.28\textwidth}{!}{\includegraphics*{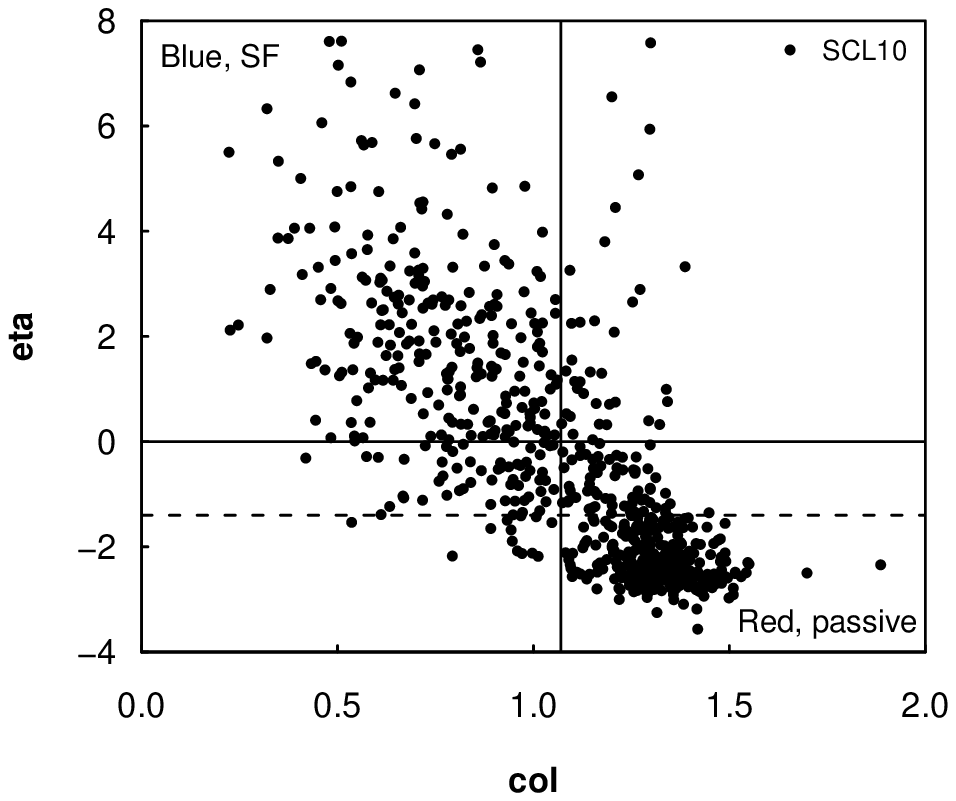}}
\caption{The spectral parameter $\eta$ and the colour index $col$ for
galaxies in rich superclusters.  The solid lines separate passive and
star-forming galaxies, the dashed line -- early and late-type galaxies.
From left to right: SCL126, SCL9, SCL10.}
\label{fig:etacol}
\end{figure*}

Next we compare the statistical properties of populations of galaxies
in superclusters.  We plot in Fig.~\ref{fig:etacol} the values of the
spectral parameter $\eta$ and the colour index $col$ of all
supercluster galaxies.  In this figure we also plot the values of
$\eta$ and $col$ used to divide galaxies into populations of early and
late type galaxies, and into passive and actively star forming
galaxies. Fig.~\ref{fig:etacol} shows that galaxies from different
superclusters populate the $\eta$--$col$ diagram in a somewhat
different manner. In the supercluster SCL126 red and passive galaxies
dominate (the lower right part of the figure). The fraction of blue
galaxies with the colour index $col < 0.6$ and the fraction of
galaxies of type 4 (the spectral parameter $\eta > 3.5$, Sec. 2.3) in
this supercluster is very small. In the supercluster SCL9 the most red
galaxies have larger colour indices $col$ and smaller spectral
parameters $\eta$ than in the supercluster SCL126 -- the reddest
galaxies in our sample reside in the supercluster SCL9. 
Furthermore, the fraction of blue galaxies with the colour index $col < 0.6$
and the fraction of galaxies of type 4 (the spectral parameter $\eta >
3.5$) in this supercluster is 
even smaller than in the supercluster SCL126.

In the supercluster SCL10 the populations of galaxies are different:
there are only a few very red, early type galaxies in this
supercluster.  The fraction of blue, late type, star forming galaxies
in this supercluster is larger than the fraction of these galaxies in
the superclusters SCL126 and SCL9.

Now let us study in more detail where galaxies of different
populations reside within superclusters. For that, we compare the
galaxy populations in rich and poor groups and those not in any
group. We divide the groups by their richness as follows: rich groups
($N_{gal} \geq 10$, $Gr_{10}$), poor groups ($N_{gal} < 10$, $Gr_2$),
and isolated galaxies (those galaxies in rich superclusters which do
not belong to any group, $IG$). We shall use the group richness as a
local density indicator, to compare the properties of galaxies in
various local environments (Paper III). To see whether the galaxy
populations depend also on the global density in superclusters, we
further divide groups by their environmental density $D1$, $D2$ and
$D3$.

\begin{figure*}[ht]
\centering
\resizebox{0.28\textwidth}{!}{\includegraphics*{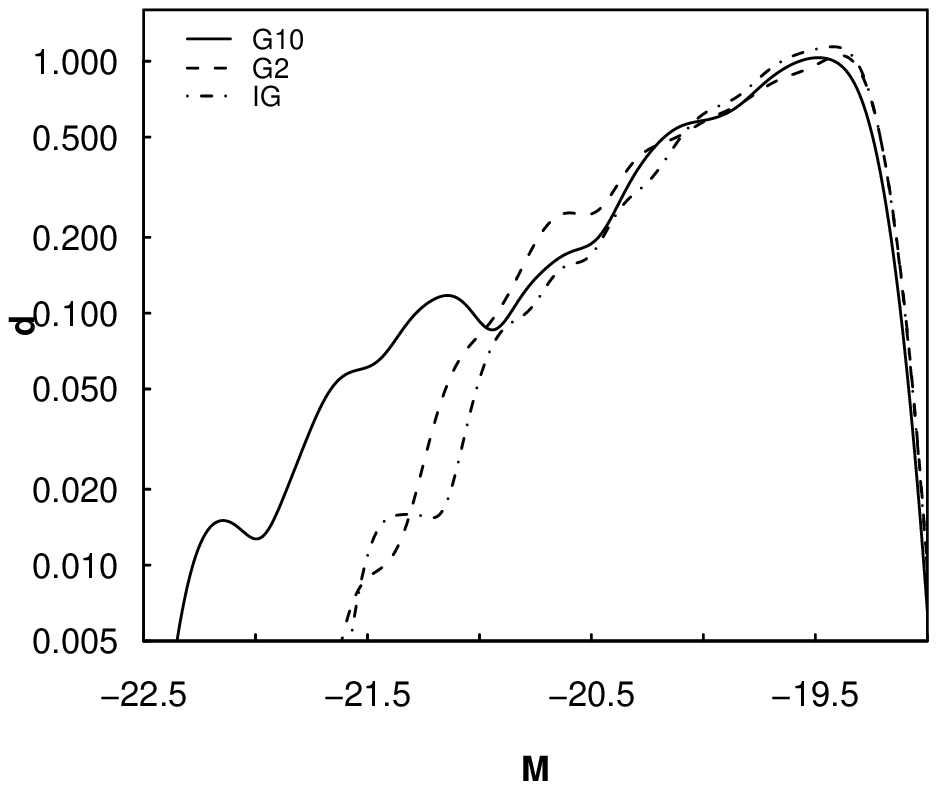}}
\resizebox{0.28\textwidth}{!}{\includegraphics*{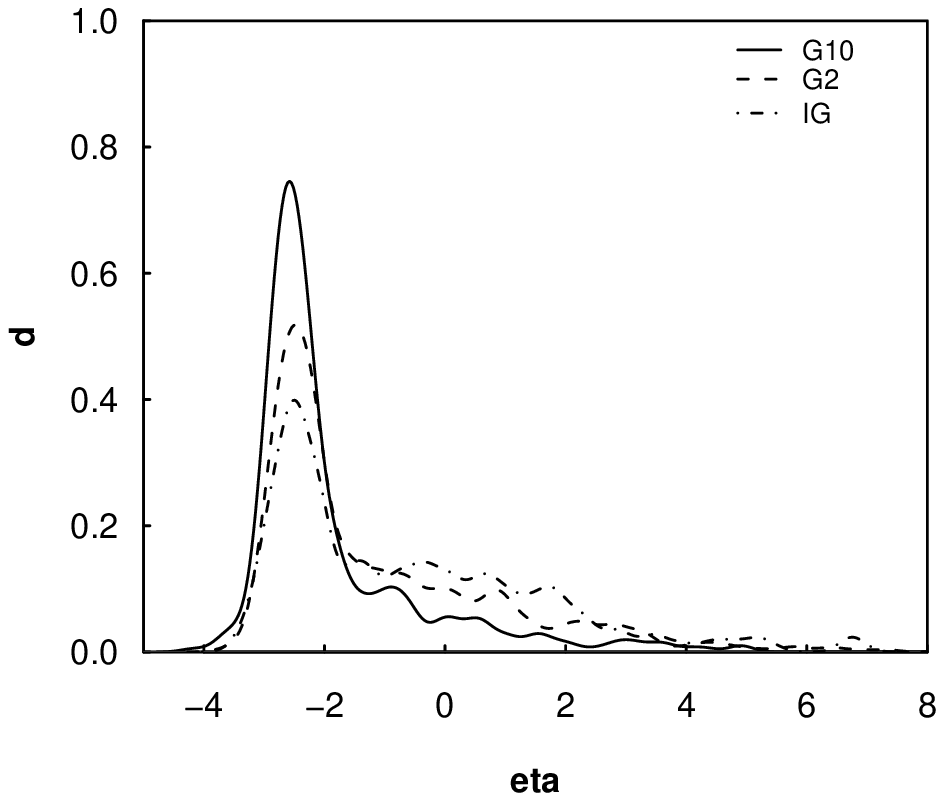}}
\resizebox{0.28\textwidth}{!}{\includegraphics*{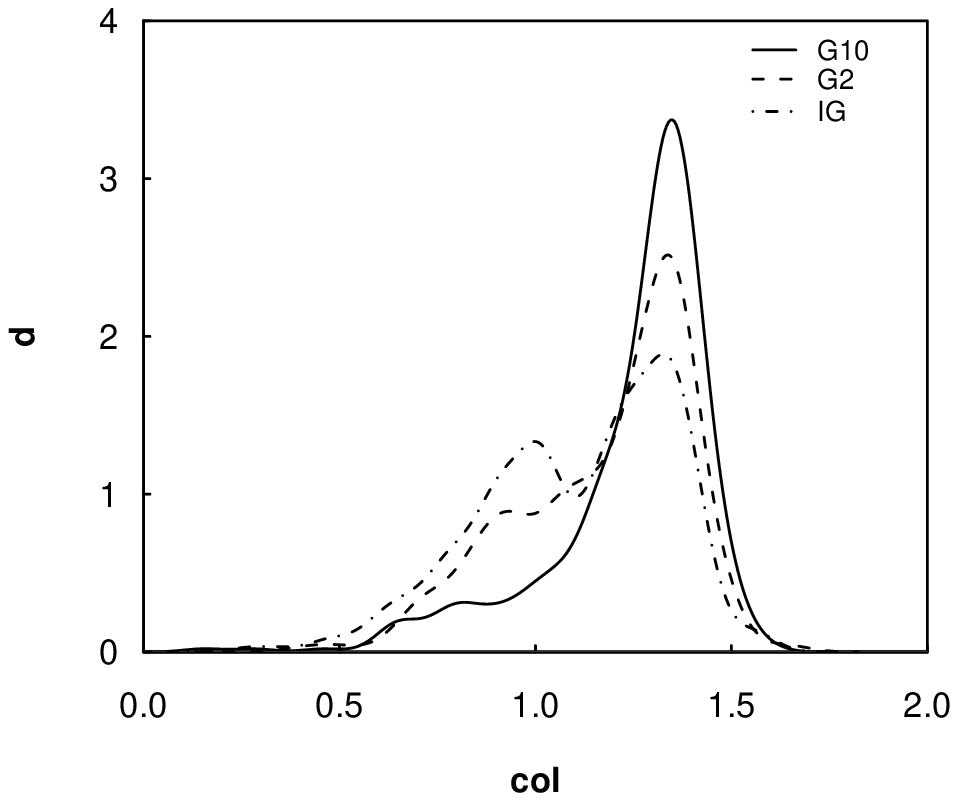}}
\hspace*{2mm}\\
\resizebox{0.28\textwidth}{!}{\includegraphics*{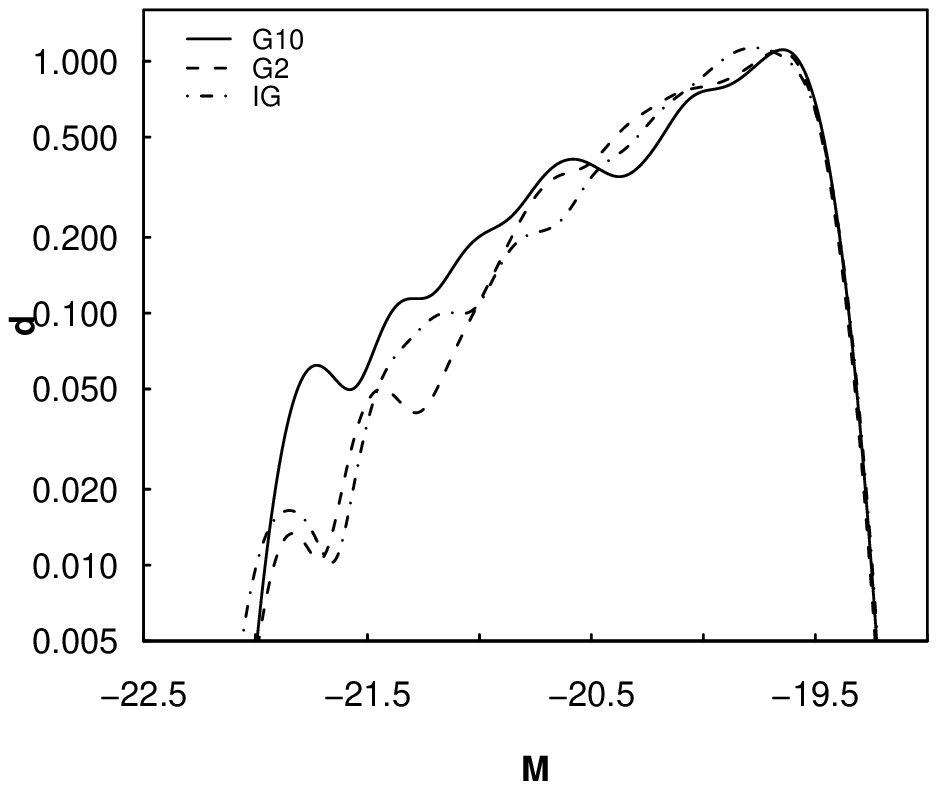}}
\resizebox{0.28\textwidth}{!}{\includegraphics*{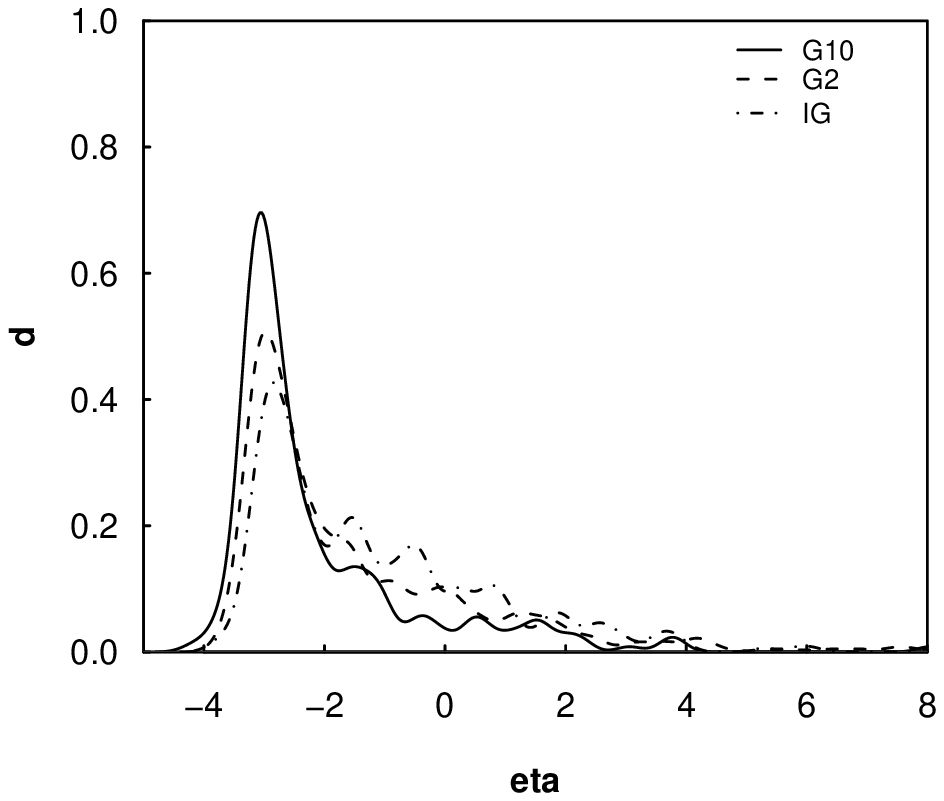}} 
\resizebox{0.28\textwidth}{!}{\includegraphics*{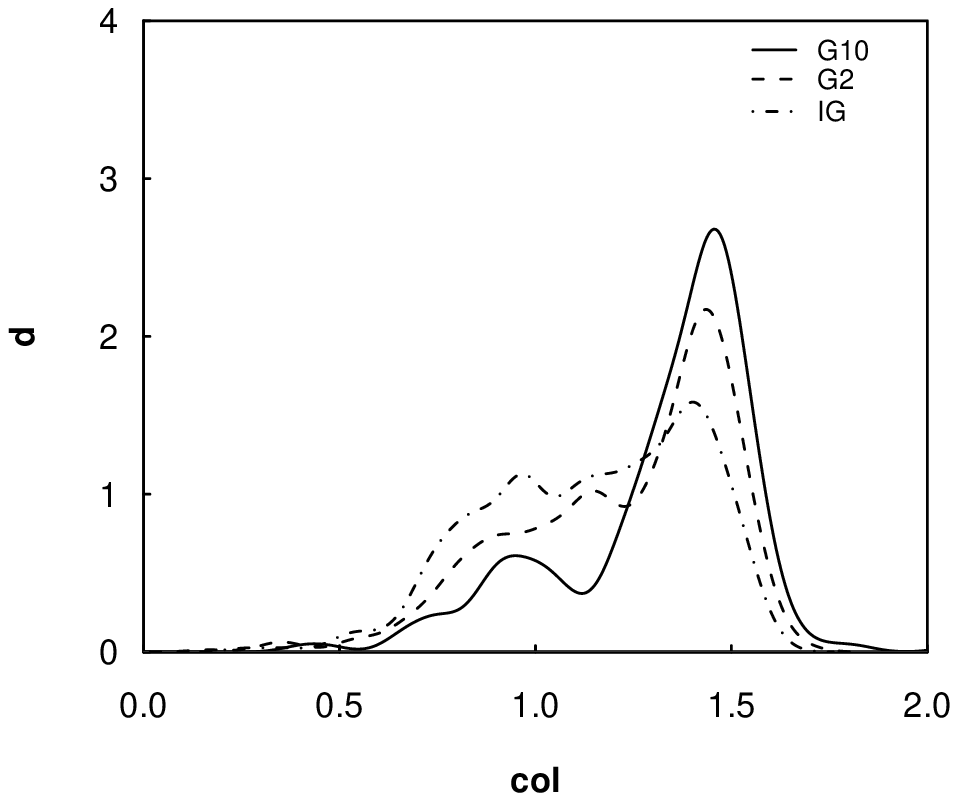}} 
\hspace*{2mm}\\
\resizebox{0.28\textwidth}{!}{\includegraphics*{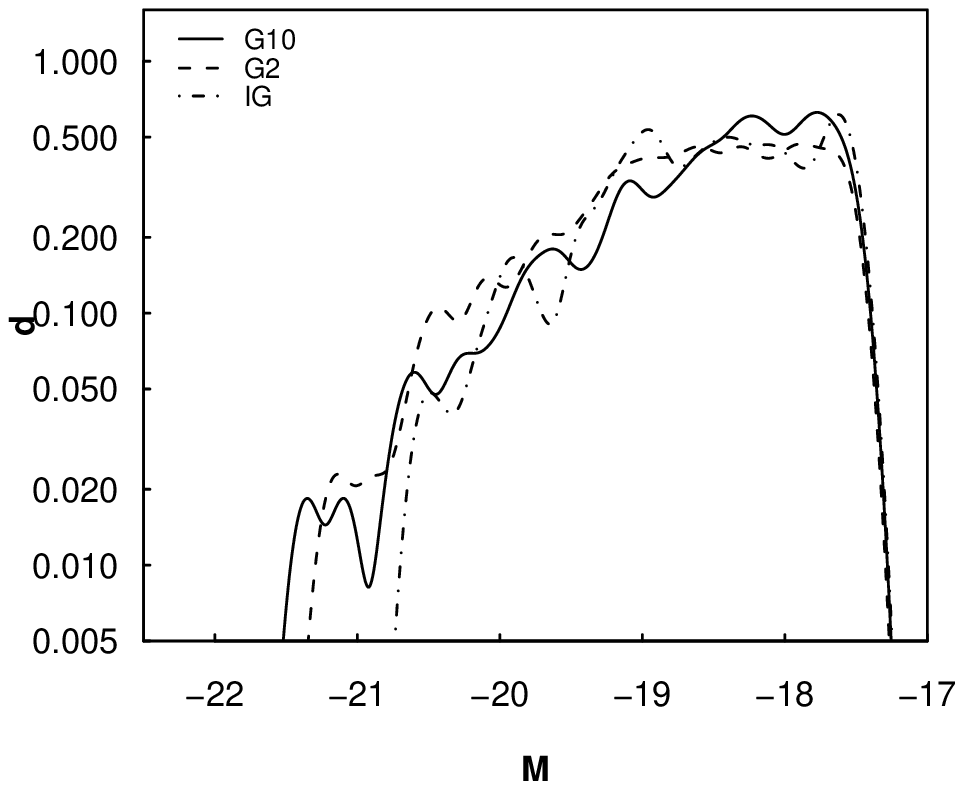}}
\resizebox{0.28\textwidth}{!}{\includegraphics*{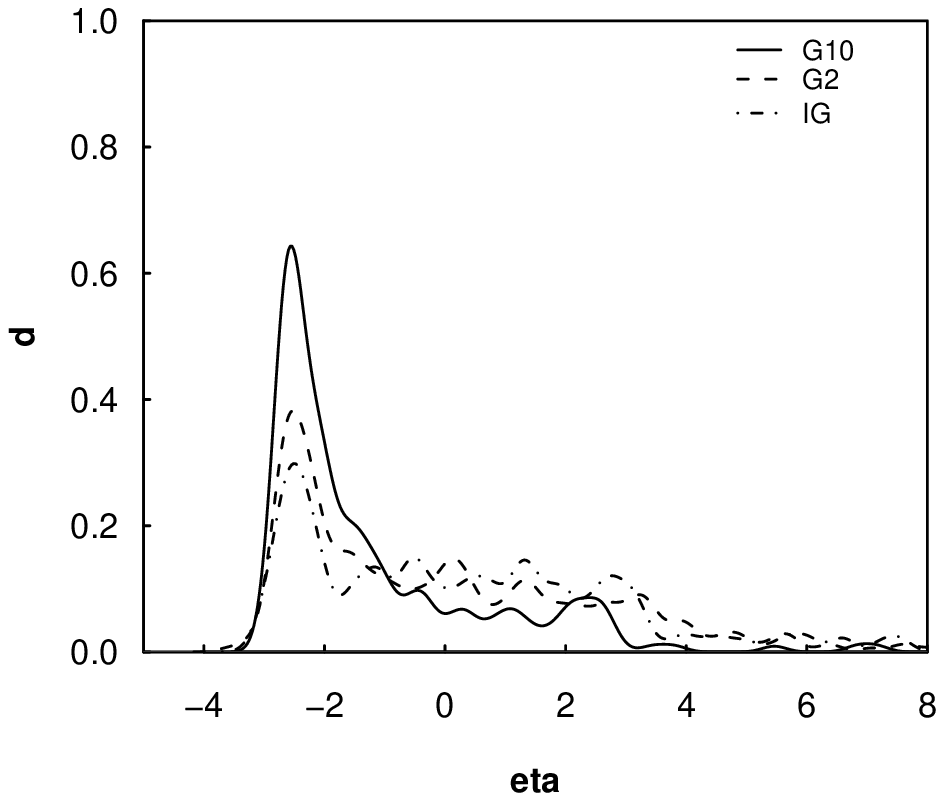}} 
\resizebox{0.28\textwidth}{!}{\includegraphics*{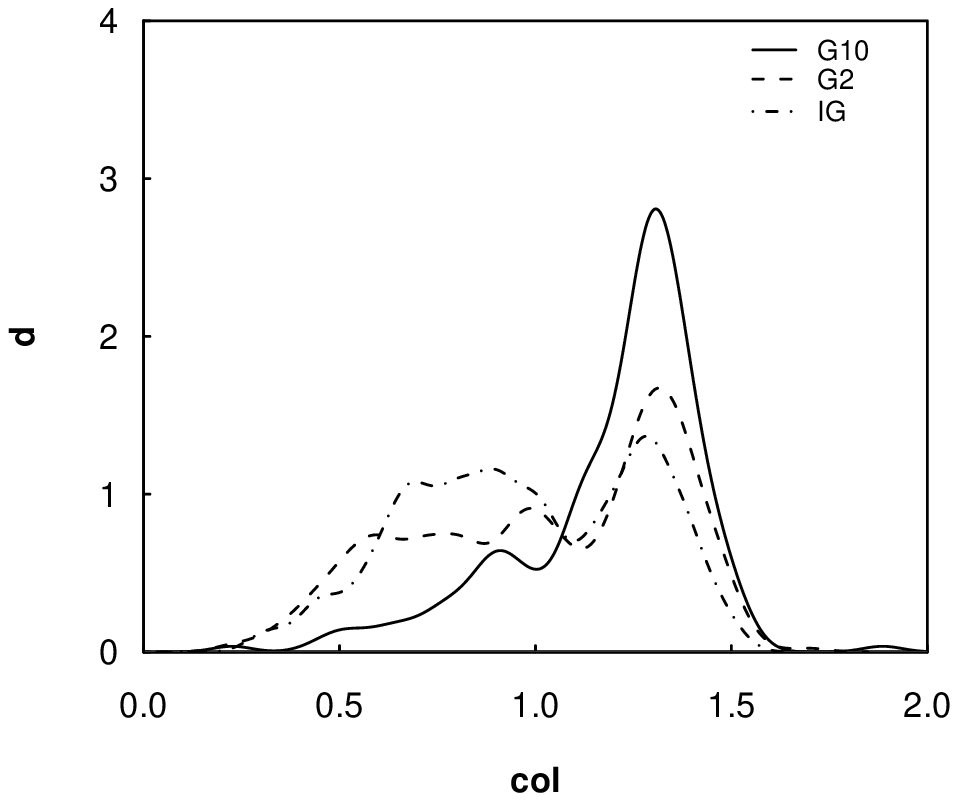}} 
\caption{The distribution of the absolute magnitude (left panels),
the spectral parameter $\eta$ (middle panels) and
 the colour index $col$ (right panels) 
of galaxies in groups of various richness
in  the supercluster SCL126 (upper row), in SCL9 (middle row)
and in SCL10 (lower row). The populations are:
$G10$) -- rich groups ($N_{gal} \geq 10$),
$G2$ -- poor groups ($N_{gal} < 10$), and $IG$ -- isolated
galaxies (those galaxies which do not belong to
any group).
}
\label{fig:sclg}
\end{figure*}

We plot the differential luminosity functions and the distributions of
the spectral parameter $\eta$ and the colour index $col$ for galaxies
in groups of different richness in Fig.~\ref{fig:sclg}. In
Table~\ref{tab:5} we present the ratios of the numbers of bright and
faint galaxies $B/F$ in superclusters, and the ratios of early and
late type galaxies $E/S$, as classified by the spectral parameter
$\eta$. We also calculate the ratio of the numbers of passive galaxies
and actively star forming galaxies $q/SF$, using the spectral
parameter $\eta$, and the ratio $P/A$, classified by the color index
$col$. In this Table we also give the ratios of the numbers of
galaxies of different type in the supercluster SCL126 for the
magnitude limit $M_{bj} = -19.50$, for comparison with SCL9.

Fig.~\ref{fig:sclg} (upper row) shows the galaxy populations in groups of 
different richness in the supercluster SCL126. We see that the brightest 
galaxies reside in rich groups. Also, there are relatively more early 
type, red galaxies in rich groups than in poor groups or among isolated 
galaxies. Both  the spectral parameter $\eta$ and the colour index $col$ 
indicate that the fraction of star-forming galaxies increases, as we 
move from higher local densities (rich groups) to lower local densities 
(poor groups and those galaxies which do not belong to groups). In 
addition, Table~\ref{tab:5} shows that the fraction of early type, red, 
passive galaxies in rich groups is the largest in the highest density 
region, $D1$, and the lowest in rich groups of the lowest density 
region, $D3$. In the high density region $D1$ among those galaxies 
which do not belong to groups, there is a larger percentage of passive, 
non-star-forming, red galaxies, than among isolated galaxies in the 
lower density regions. 

In other words, the whole high-density core of this supercluster contains 
relatively more passive, red, early-type galaxies than the lower-density 
region, where there are more blue, star-forming 
galaxies. This shows that both the local (group/cluster) and the global 
(supercluster) environments are important in influencing types, colours 
and star formation rates of galaxies.

Earlier studies (Davis \& Geller \cite{dg76}; Dressler \cite {d80}; 
Phillips et al. \cite{phill98}; Norberg et al. \cite{n01}, \cite{n02}; 
Zehavi et al. \cite{zehavi02}; Goto et al. \cite{go03}; Hogg et al. 
\cite{hogg03}, \cite{hogg04}; Balogh et al. \cite{balogh04}; De Propris 
et al.  \cite{depr03}; De Propris et al. \cite{depr04}; Madgwick et al. 
\cite{ma03b}; Croton et al. \cite{cr05}; Blanton et al. \cite{blant04}, 
\cite{blant06} among others) have shown 
the difference between galaxy populations in clusters and in the field. 
Our calculations show that this difference exists also between the core 
regions and the outer regions of rich superclusters (see also Paper III).

Now let us study the distribution of galaxies from different populations in
the supercluster SCL9 (Fig.~\ref{fig:sclg}, middle row).  We see, in
accordance with the data about SCL126, that the fraction of the brightest
galaxies, as well as the fraction of early type, red galaxies is the largest
in rich groups and the smallest among those galaxies which do not belong to
groups. However, when we compare the galaxy content of groups from this
supercluster in regions of different density, $D1$--$D3$ (Table~\ref{tab:5}),
we see that in this supercluster the fraction 
of early type, red galaxies is
the largest in the region of intermediate density, $D2$. This may be due to the
presence of X-ray clusters in this region, which affect 
the galaxy content 
(see also Hilton et al. \cite{hil05} and Paper III).  Also, since the 2dF
survey covers only a part of this supercluster, the density in this region may
be underestimated.

However, the most interesting feature seen in this Figure and in
Table~\ref{tab:5} is the striking difference between the galaxy
properties in the superclusters SCL126 and SCL9. In Table~\ref{tab:5} we
gave the ratios of the numbers of galaxies of different type in the
supercluster SCL126 also for the magnitude limit $M_{bj} = -19.50$, in
order to compare these ratios with those calculated for the
supercluster SCL9. Our calculations show that in the supercluster
SCL126 the ratio of the numbers of red and blue galaxies for the
magnitude limit $M_{bj} = -19.50$ is 2.77, while for the same
magnitude limit in SCL 9 this value is 2.45.  The fraction of red
galaxies in SCL126 is the largest in rich groups -- 7.37 in the region
$D1$, while in the region $D2$ of SCL9 this value is 6.00, which is
the largest value in SCL9.  The differences between the fractions of the
numbers of early and late type galaxies, as well as between the
fractions of quiescent and actively star forming galaxies (classified
by the spectral parameter $\eta$) in the superclusters SCL126 and SCL9
are smaller; these fractions are even slightly larger in SCL9 than in
SCL126, but in the rich groups of SCL126 the fraction of both early
type galaxies and quiescent galaxies is much larger than in the rich
groups of SCL9.

Fig.~\ref{fig:sclg} demonstrates that differences between galaxy 
populations of the superclusters SCL126 and SCL9 are even larger than 
the ratios in Table~\ref{tab:5} show. In SCL9 there are more blue 
galaxies than in SCL126. This difference in seen among the galaxies in 
groups and among those galaxies which do not belong to groups. 
Furthermore, in SCL9 there is a small fraction of extremely red galaxies 
which are absent in SCL126. It is difficult to ascribe these differences 
between galaxy populations to selection effects due to the different 
distances of these superclusters -- in that case we should expect that 
in the supercluster SCL126 the fraction of early type, passive galaxies 
should be smaller than in SCL9, which is opposite to our results.

The brightest galaxies in SCL126 have larger luminosities than those
in SCL9.  
Also, in SCL126 there is a large difference
between the luminosities of the brightest galaxies in rich and poor
groups, this difference in SCL9 is much smaller.

The populations of galaxies in the supercluster SCL10 are again
somewhat different. There are more late type spiral and irregular
galaxies here (with large values of $\eta$), and more blue galaxies
(with small values of $col$), both among the galaxies in groups and
among isolated galaxies (especially in regions of lower global
density) in this supercluster than in the other two rich superclusters
studied in this paper.  In this supercluster, the luminosity of the
brightest galaxies in rich and poor groups is similar, but, overall,
the luminosities of the brightest galaxies in this supercluster are
lower than the luminosities of the brightest galaxies in SCL126 and
SCL9.

One reason of the differences between the galaxy populations of the
supercluster SCL10 and other superclusters is, of course, the influence of the
lower luminosity limit used for this supercluster in our study.  However, this
does not explain why there are no very luminous galaxies, and also no very red
galaxies in this supercluster (Fig.~\ref{fig:etacol}). The reason for this may
be the following: in the case of SCL10 most 
of its rich (Abell) clusters lie
outside of the 2dF survey volume (Table 1). As the reddest, early type, very
bright galaxies reside in rich clusters, the reason why we do not see these
galaxies may be that these rich clusters lie outside the survey limits. Thus,
in the case of this supercluster, we actually study the populations of
galaxies located in the outer parts of the supercluster.  These galaxies are
fainter, thus 
due to the luminosity limits, we see in other superclusters a
smaller fraction of these galaxies.

Our data show that galaxy populations in superclusters depend both on
the local and global densities.  For example, even in rich groups the
fraction of blue, star-forming galaxies in the region $D3$ is larger
than in the region $D1$. The same can be said for the spectral
parameter $\eta$ -- the fraction of late-type galaxies (types 2--4) in
the region $D3$ is larger than in the region $D1$. The
Kolmogorov-Smirnov test shows that the differences between the
distributions of colour indices of galaxies from regions of different
density in all superclusters (with two exceptions) are statistically
significant at least at the 99\% confidence level. In the supercluster
SCL9 these differences between the galaxies from the density regions
$D1$ and $D2$, and in the supercluster SCL10 from the density regions
$D2$ and $D3$ are statistically significant at least at the 87\%
confidence level.

\section{Discussion}

\subsection{Comparison with other very rich superclusters}

The richest relatively nearby superclusters are the Shapley Supercluster
(Proust et al. \cite{proust06}, Bardelli et al. \cite{bar00}, 
Quintana, Carraso and Reisenegger \cite{qui00} and references therein) and
the Horologium--Reticulum Supercluster (Rose et al. \cite{rose02}, 
Fleenor et al. \cite{fleenor05}, Einasto et al. \cite{e03d}).

The Shapley Supercluster contains 28 Abell clusters, 8 X-ray clusters
among them, and one non-Abell X-ray cluster according to the catalogue
by Einasto et al. (\cite{e2001}). Proust et al. (\cite{proust06}) list
over 40 rich clusters of galaxies in the redshift range of the Shapley
supercluster, and an overdensity of about 5.4.  The spatial extent of
this supercluster is over 100~\Mpc\ and Proust et al. estimate that
the total mass of the Shapley supercluster is at least $M_{tot} = 5\times
10^{16} h^{-1} M_{\sun}$.  The main core of this supercluster contains at
least two Abell clusters and two X-ray groups, and the overdensity of
galaxies in the central region is at least 11 (Bardelli et
al. \cite{bar00}). The differences between the various density
estimates are due to different methods used, and to different
scales at which overdensities are estimated.

Haines et al. (\cite{hai06}) demonstrated that the colours of galaxies 
in the core region of the Shapley supercluster depend on their 
environment, with redder galaxies being located in clusters.  They also 
found a large amount of faint blue galaxies between the clusters. 

Another prominent supercluster is the Horologium--Reticulum
supercluster in the Southern Sky.  This supercluster contains 35 Abell
clusters, 10 X-ray clusters among them, and one non-Abell X-ray
cluster (Einasto et al. \cite{e2001}). This supercluster is a dominant
structure in all the Southern slices of the Las Campanas Redshift
Survey (Einasto et al. \cite{e03d}). The Horologium--Reticulum
supercluster has been studied in detail by Rose et al. (\cite{rose02})
and Fleenor et al. (\cite{fleenor05}). This supercluster contains
three concentrations of clusters, groups and galaxies, which are
connected by filaments of galaxies and groups.  The spatial extent of
this supercluster is over 100~\Mpc, and the mean overdensity of
galaxies is about 3 (Fleenor et al. (\cite{fleenor05}). Fleenor et al.
estimated that the total mass of the Horologium--Reticulum
supercluster may be of the same order as the total mass of the Shapley
supercluster.

The richest superclusters in our sample are the superclusters SCL126
and SCL9.  Of these two, the supercluster SCL9 (the Sculptor
supercluster) is comparable to the Shapley and the
Horologium-Reticulum superclusters, 
both by their size and by the number
of rich clusters in it. In the Sculptor supercluster there are 6 X-ray
clusters. In addition, Zappacosta et al. (\cite{zap04})
cross-correlated ROSAT observations with the galaxy density map in this
supercluster and found evidence about the presence of warm-hot diffuse
gas which is associated with the intercluster galaxy distribution.

If we assume that the mean mass-to-light ratio is about 400 (in solar
units), then we can estimate the masses of the richest superclusters
of our sample, using estimates of the total luminosity of the
superclusters (Table~\ref{tab:1}): $M_{SCL9} = 2 \times 10^{16}  h^{-1} 
M_{\sun}$, $M_{SCL126} = 1.5 \times 10^{16}  h^{-1}  M_{\sun}$.  These
estimates are lower limits only, since the 2dF survey does not fully
cover these superclusters.

The supercluster SCL10 was recently studied by Porter and Raychaudhury
(\cite{pr05}), using data about Abell clusters, galaxy groups (by Eke
et al. \cite{eke04}) and galaxies in this supercluster. They estimate
that the total mass of the Pisces-Cetus supercluster is at least
$M_{tot} = 1.5 \times 10^{16} h^{-1}  M_{\sun}$; the spatial extent of this
supercluster is more than 100\Mpc\ (Jaaniste et al. \cite{ja98}).
Porter and Raychaudhury determined also the star formation rates for
galaxies in groups in this supercluster, 
according to the spectral
index $\eta$. They concluded that galaxies in rich clusters have lower
star formation rates than galaxies in poor groups in agreement with
our results. In another paper Porter and Raychaudhury 
demonstrated  that in the filament between the clusters in this
supercluster the fraction of star forming galaxies is higher at larger
distance from clusters than close to clusters (Porter and Raychaudhury
\cite{pr06}). 
At the same time our
analysis shows that in the regions $D2$ and $D3$ of this supercluster
the fraction of actively star forming galaxies among those galaxies
which do not belong to groups is the highest among the superclusters
studied in this paper. This is probably a reflection of the same result
as obtained by Porter and Raychaudhury.

Gray et al. (\cite{gray04}) studied the supercluster A901/902 and detected
strong evidence that the highest density regions in clusters are populated
mostly with quiescent galaxies, while star forming galaxies dominate in
outer/lower density regions of clusters.

Our present data reveals the dependency of the properties of galaxies in 
superclusters on both the local density (as shown also by the 2DF team) 
and on the global density (see also Paper III and references therein).

\subsection{Properties of groups in superclusters}

We showed that groups in high-density cores of superclusters are richer
than groups in low-density regions of superclusters.  In high-density
cores of superclusters, groups contain relatively more passive, red,
early-type galaxies than the groups in lower-density regions in
superclusters, where there are relatively more blue, star-forming
galaxies. Therefore, both the richness of a group and its galaxy content
depend on the large scale environment where it resides.

Plionis (\cite{pl04} and references therein) showed that the dynamical
status of groups and clusters of galaxies depends on their large-scale
environment. Here we show that also the richness of a group and its galaxy
content depend on the large-scale environment. The first of these effects
was described as an environmental enhancement 
of group's richness in Einasto et
al. (\cite{e03c}) and (\cite{e03d}).

\subsection{Substructures and galaxy populations in superclusters --
evidence about differences of their formation and evolution?}

We showed that there exist several differences between the properties 
of individual rich superclusters, which cannot be explained by selection 
effects only. The overall structure of the supercluster SCL126 resembles 
a rich filament with several branches and a smaller number of substructures 
than in the supercluster SCL9, which resembles a multispider having a 
number of clumps connected by relatively thin filaments. 

The supercluster SCL126 has a very high density core with several Abell 
and X-ray clusters in a region with dimensions less than 10 \Mpc\ (see 
also Einasto et al. \cite{e03d}). Such very high densities of galaxies 
have been observed so far only in a very few superclusters. Among them 
are the Shapley supercluster (Bardelli et al. \cite{bar00}), the 
Aquarius supercluster (Caretta et al. \cite{car02}), and the Corona 
Borealis supercluster (Small et al. \cite{small98}). A very small number 
of such a high density cores of superclusters is consistent with the 
results of N-body simulations which show that such high density regions 
(the cores of superclusters that may have started the collapse very 
early) are rare (Gramann \& Suhhonenko \cite{gra02}). The fraction of 
early type, passive, non-star-forming galaxies in the core region of 
the supercluster SCL126 is very high, both in groups and among those 
galaxies which do not belong to groups. This indicates that the 
properties of galaxies and their evolution history have been affected by 
both local and global densities in superclusters.
  
In the supercluster SCL126 the fraction of blue, actively star forming 
galaxies is smaller than in the supercluster SCL9, especially in its 
high density core. This, together with the presence of a high density 
core and overall more homogeneous structure than in the supercluster 
SCL9 may be an indication that the supercluster SCL126 has  formed 
earlier (is more evolved dynamically by the present epoch)
than the supercluster SCL9.

\section{Conclusions}

We used a catalogue of superclusters of galaxies for the 2dF Galaxy
Redshift Survey  to study the internal structure and galaxy populations
of the richest superclusters.  Our main conclusions are the following.

\begin{itemize}

\item{} 
We study substructures in superclusters using their detailed density 
distributions and the fourth Minkowski functional, $V_3$. We show that
substructures in different superclusters differ strongly. Large differences 
are seen also in a way how substructures are delineated by 
galaxies from different populations.

\item{}
The values of the fourth Minkowski Functional $V_3$, 
which contain information about both local and global morphology, show   
the clumpiness of  superclusters as determined by galaxies from
different populations. 

\item{}
In the supercluster SCL126 the number of clumps
in the distribution of red, passive 
galaxies is larger than the number of clumps determined by blue,
star-forming galaxies -- 
such galaxies surround  the high-density regions
rather uniformly. In the supercluster SCL9 the values of $V_3$
are different, indicating that star-forming galaxies are located in
numerous small clumps around passive galaxies, which are located
preferentially in rich groups and clusters.

\item{} 
Groups in high-density cores of superclusters are richer than
in lower-density (outer) regions of superclusters. In high-density
cores, groups contain relatively more passive, red, early-type
galaxies than the groups in lower-density regions in superclusters, where
there are more blue, star-forming galaxies.  Therefore, both the
richness of a group and its galaxy content depend on the 
large scale environment where it resides.

\item{}
In the high density cores of all rich 
superclusters, even among those galaxies which do not belong to galaxy 
clusters, there is a large fraction of passive, non-star-forming, red 
galaxies. Among the isolated galaxies in the lower density supercluster 
regions, the red galaxy fraction is much lower. 

\item{}
Unexpectedly, we found that there are 
several differences between the galaxy populations of different superclusters,
which cannot be explained by selection effects only. 
In particular, the
supercluster SCL126 contains a larger fraction of early type, 
red, passive galaxies than the supercluster SCL9, especially 
%vm
in its high density core region.

\item{}
In the supercluster SCL9 there is a population of very red and early 
type galaxies, which reside in the richest groups of region of high (but 
not the highest) global density.  Such very red galaxies are absent in 
SCL 126.

\item{}
In SCL126, the most luminous galaxies in rich groups have much
larger luminosities than most luminous galaxies in poor groups, 
while in SCL9 and SCL10 the luminosity of the brightest galaxies in rich 
and poor groups is comparable. 

\item{}
The presence of a high density core with X-ray clusters and a relatively 
small fraction of star-forming galaxies in the supercluster SCL126 may 
be an indication that this supercluster has been formed earlier 
(is more advanced dynamically) than SCL9. 

\item{}
In SCL10 there are no very bright galaxies, as well as no very red 
and very early type galaxies. Also, there are relatively more
star-forming, late type, blue galaxies than in the larger superclusters
SCL126 and SCL9. This
may be partly due to selection effects, but partly it may be  due to the fact 
that the 2dF survey, our base data sample, does not cover
this supercluster in full.
\end{itemize}

Our study indicates the importance of the role of superclusters
as a high density environment, which affects the properties 
of their member galaxies and the groups/clusters of galaxies
that constitute the supercluster.

The forthcoming Planck satellite observations will determine the
anisotropy of the cosmic background radiation with unprecedented
accuracy and angular resolution. As a by-product, Planck measurements
will provide an all-sky survey of massive clusters via the
Sunyaev-Zeldovich (SZ) effect. For the Planck project, detailed
information of supercluster properties is important, helping to
minimize anomalies in CMB maps at low multipoles. These anomalies are
partly caused by imprints of local superclusters, especially of their
cores, on the SZ-signals. As continuation of the present work, we are
preparing supercluster catalogues for the Planck community.

\begin{acknowledgements}

  We are pleased to thank the 2dFGRS Team for the publicly available 
  data releases.  We thank T\~onu Viik for helpful suggestions. The 
  present study was supported by the Estonian Science Foundation grants 
  No. 6104 and 7146, and by the Estonian Ministry for Education and 
  Science research project TO 0060058S98. This work has also been 
  supported by the University of Valencia through a visiting 
  professorship for Enn Saar and by the Spanish MCyT project AYA2003-
  08739-C02-01 (including FEDER).  J.E.  thanks Astrophysikalisches 
  Institut Potsdam (using DFG-grant 436 EST 17/4/06),  and the Aspen 
  Center for Physics for hospitality,  where part of this study was 
  performed.   
PH and PN were supported by Planck science in Mets\"ahovi, Academy of Finland.
  In this paper we have used R, a language for data
  analysis and graphics (Ihaka \& Gentleman \cite{ig96}).

\end{acknowledgements}

\end{document}